\documentclass[letterpaper,twocolumn,10pt]{article}
\usepackage[breaklinks,backref=page]{hyperref}
\usepackage{style/usenix}

\usepackage{subcaption}
\expandafter\def\csname ver@subfig.sty\endcsname{}

\usepackage{microtype}
\usepackage{tikz}
\usepackage{subfig}
\usepackage{multirow}
\usepackage{booktabs}
\usepackage{graphicx}
\usepackage{xspace}
\usepackage{algorithm}
\usepackage[noend]{algpseudocode}
\usepackage{balance}
\usepackage{amsmath}
\usepackage{mathtools}
\usepackage{amsfonts}
\usepackage{amssymb}
\usepackage{amsthm}
\usepackage{threeparttable}
\usepackage{adjustbox}
\usepackage[available, functional]{style/usenixbadges}

\usepackage{xcolor}
\usepackage{enumitem,kantlipsum}
\usepackage{siunitx}
\usepackage[compact]{titlesec}
\usepackage[nosort]{cleveref} % cleverref must be kept last: https://tex.stackexchange.com/questions/148699/equation-reference-undefined-when-using-cref-and-packageamsmath
\usepackage{url}
\usepackage{breakurl}

\newcommand{\argmax}{\operatornamewithlimits{arg\,max}}

\newtheorem{theorem}{Theorem}
\newtheorem{lemma}{Lemma}
\renewcommand{\thefootnote}{\arabic{footnote}}
\newcommand{\Sys}{{FlexLLM}\xspace}
\newcommand{\sys}{{FlexLLM}\xspace}
\newcommand{\llama}{LLaMA\xspace}

\newcommand{\commentout}[1]{}
\algnewcommand{\LeftComment}[1]{\Statex \(\triangleright\) #1}

\newcommand{\m}[1]{\mathcal{#1}}
\newcommand{\removed}[1]{}

\newcommand{\peft}{\textproc{PEFT}\xspace}


\newif\iffinal
\finalfalse

\newcommand*{\cmu}{\includegraphics[scale=0.0018]{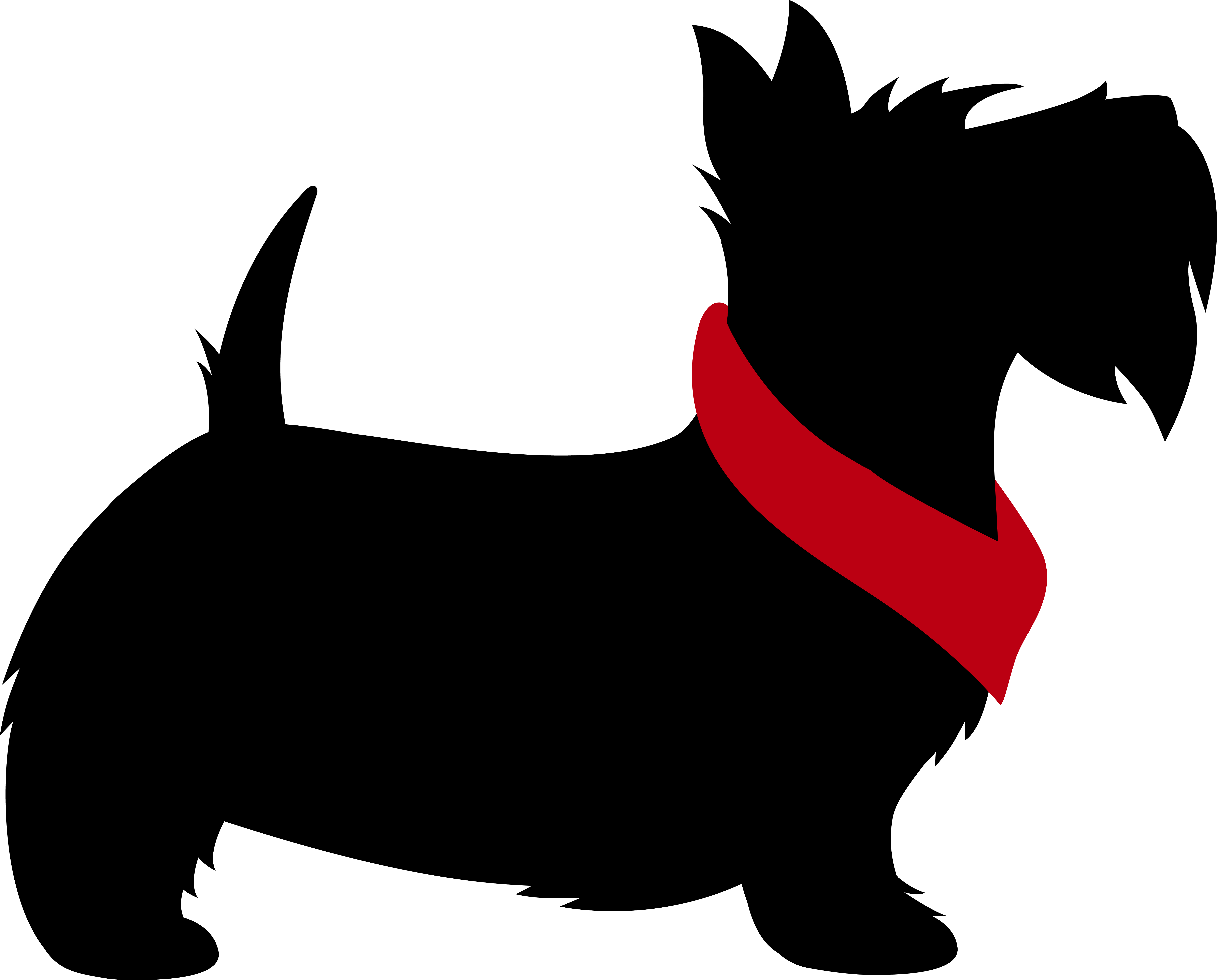}}
\newcommand*{\purdue}{\includegraphics[scale=0.025]{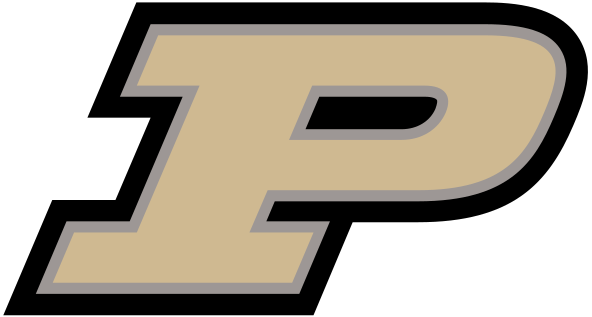}}
\newcommand*{\anthropic}{\includegraphics[scale=0.06]{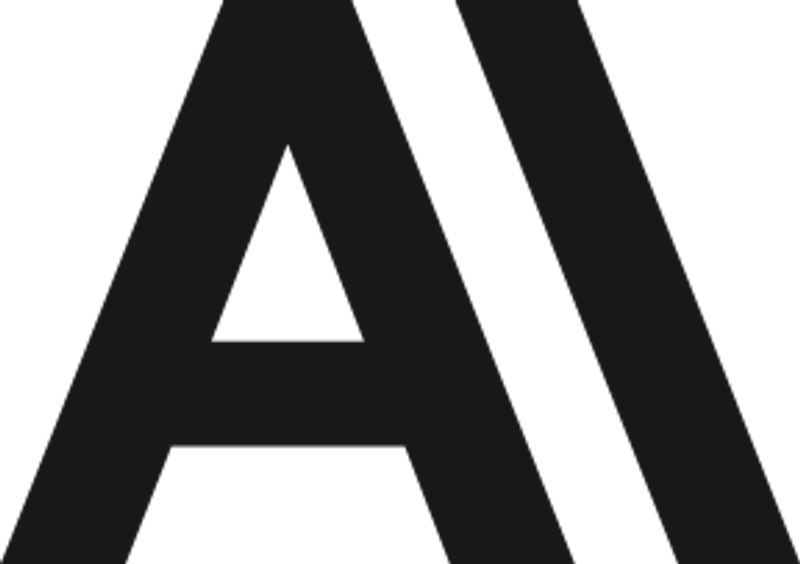}}
\newcommand*{\stanford}{\includegraphics[scale=0.035]{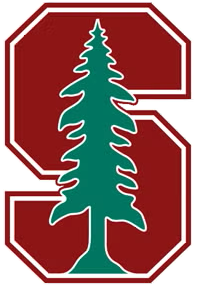}}
\newcommand*{\mistral}{\includegraphics[scale=0.06]{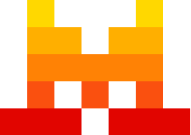}}
\newcommand*{\aws}{\includegraphics[scale=0.025]{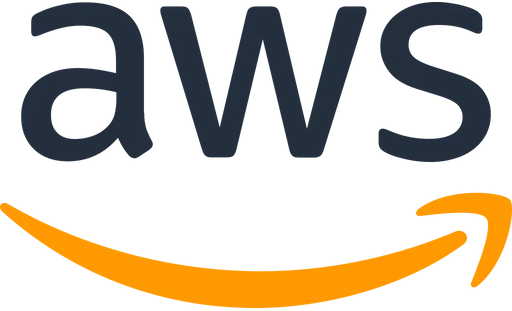}}

\begin{document}

\title{\sys: Token-Level Co-Serving of LLM Inference\\and Finetuning with SLO Guarantees}

\author{
\rm
Gabriele Oliaro$^{*}$ \cmu \quad
Xupeng Miao$^{*\dag}$ \purdue \quad
\\
\rm
Xinhao Cheng \cmu \quad
Vineeth Kada$^{\dag}$ \anthropic \quad
Mengdi Wu \cmu \quad
Ruohan Gao \cmu \quad
Yingyi Huang \cmu \quad
\\
\rm
Remi Delacourt$^{\dag}$ \mistral \quad
April Yang \cmu \quad
Yingcheng Wang$^{\dag}$ \purdue \quad
Colin Unger \stanford \quad
Zhihao Jia \cmu \,\aws
\\ \\ %[.3em]
\cmu \; Carnegie Mellon University \quad
\purdue \; Purdue University \quad
\anthropic \; Anthropic PBC\\
\stanford \; Stanford University \quad
\mistral \; Mistral AI \quad
\aws \; Amazon Web Services
}

\maketitle

{\let\thefootnote\relax\footnotetext{$^*$Equal contribution.}}
{\let\thefootnote\relax\footnotetext{$^{\dag}$Work done at CMU.}}

\pagestyle{plain}

\begin{abstract}
    Finetuning large language models (LLMs) is essential for task adaptation, yet today’s serving stacks isolate inference and finetuning on separate GPU clusters—wasting resources and under-utilizing hardware. We introduce \sys, the first system to {\em co-serve} LLM inference and PEFT-based finetuning on shared GPUs by fusing computation at the token level. \sys’s static compilation optimizations—{\em dependent parallelization} and {\em graph pruning} significantly shrink activation memory, leading to end-to-end GPU memory savings by up to 80\%. At runtime, a novel {\em token-level finetuning} mechanism paired with a hybrid token scheduler dynamically interleaves inference and training tokens within each co-serving iteration, meeting strict latency SLOs while maximizing utilization. In end-to-end benchmarks on LLaMA-3.1-8B, Qwen-2.5-14B, and Qwen-2.5-32B, \sys maintains inference SLO compliance at up to 20 req/s, and improves finetuning throughput by $1.9-4.8\times$ under heavy inference workloads and $2.5-6.8\times$ under light loads, preserving over 76\% of peak finetuning progress even at peak demand. \sys is publicly available at \url{https://flexllm.github.io}.
\end{abstract}

\section{Introduction}
\label{sec:intro}
Recent advancements in large language models (LLMs) such as GPT-5~\cite{gpt5,openai2023gpt4}, DeepSeek-R1~\cite{guo2025deepseek}, LLaMA-4~\cite{touvron2023llama, llama3,llama4} and QWen 3~\cite{qwen2025qwen25technicalreport,yang2025qwen3technicalreport}, have shown strong capabilities of generating natural language texts across various application domains~\cite{zhang2019pretraining, liu2021makes, stiennon2020learning,dosovitskiy2020image}.
Given the increasingly high cost of LLM pre-training, there is a trend of standardizing foundational pre-trained LLMs and sharing sub-models across multiple downstream tasks~\cite{houlsby2019parameter,zhang2021share} through finetuning (i.e., continuous training on small task-specific datasets). For example, a series of {\em parameter-efficient finetuning} (\peft) techniques~\cite{houlsby2019parameter,li2021prefix,pfeiffer2020adapterfusion,hu2021lora,he2021towards,lester2021power} have been proposed to update a subset of trainable parameters from an LLM or introduce a small number of new parameters into the LLM while keeping the vast majority of the original LLM parameters frozen. The base LLM sharing paradigm provides opportunities to multiplex resources and improve cluster utilization (e.g., Google~\cite{barham2022pathways}, Microsoft~\cite{hu2021lora}, and Tencent~\cite{nie2023angel}).
Many recent efforts~\cite{zhou2022pets,wu2024dlora,sheng2023s,chen2023punica,zhao2024lora,iliakopoulou2024chameleon} inspired by these opportunities to build multi-task serving systems, handling inference requests for multiple fine-tuned LLMs simultaneously in a single GPU cluster.

In addition to inference, almost all LLM serving companies (including OpenAI GPT-4o~\cite{openaifinetuning}, Google Gemini~\cite{googlefinetuning}, and Databricks Mosaic AI~\cite{databricksfinetuning}), now offer commercial finetuning APIs to facilitate the creation of custom models.
This trend is not only due to the widespread adoption of LLMs but also driven by the growing demand for task-specific model customization~\cite{liu2024customizing}, data access control~\cite{tan2024democratizing}, and the need for continuous and frequent model updating~\cite{gao2023unified}.

Currently, the most popular way to offer both inference and finetuning services is to run each task on separate, {\em dedicated} clusters~\cite{weng2022mlaas,gao2022deep,chen2023deepboot}, which exclusively use hardware resources through specialized systems optimized primarily for latency and throughput, respectively. While straightforward, this practice leads to significant economic inefficiencies. In particular, resources must be overprovisioned for inference tasks, as real-world LLM inference workloads exhibit highly unpredictable, bursty request arrival patterns~\cite{wang2024towards}. For example, public reports from Microsoft and Alibaba cite average GPU utilization rates of just 52\%~\cite{lithos2025} and 10\%~\cite{lithos2025}, respectively, with production services showing similar underutilization patterns. While inference tasks must be provisioned for peak burst capacity to maintain SLOs, the resulting idle resources during normal operation cannot be utilized by finetuning tasks, which could tolerate the latency variations (minutes to hours vs. milliseconds). This is economically unsustainable given the high cost and power demands of modern GPUs, which can now exceed 1,000W per chip~\cite{lithos2025}.

In this paper, we explore a key research question: {\em Can we design a system that {\em simultaneously} serves both inference and finetuning tasks within a shared GPU cluster, dynamically adapting to unpredictable burst patterns while maintaining strict inference SLOs?}
Previous attempts (\Cref{sec:coserving}) at multiplexing inference and finetuning tasks, while successful in increasing the GPU utilization, have seen limited adoption due to their impracticality.

To solve this fundamental challenge, we introduce \textit{co-serving}, a novel multiplexing technique that effectively handles bursty workloads while satisfying strict SLOs. We built \sys, the first system for LLM inference and finetuning that implements this technique. \sys introduces a \textit{PEFT-as-a-service (PaaS)} interface, which unifies inference and finetuning tasks, enabling their joint execution on shared GPU resources.
The key insight behind co-serving is that inference and finetuning tasks, when using the same base LLMs, can be merged at fine granularity—at the level of individual tokens rather than entire requests or kernels. This token-level scheduling enables millisecond-scale resource reallocation in response to inference bursts.

This co-serving approach offers critical advantages for handling bursty workloads. When inference requests suddenly spike, \sys can instantly throttle finetuning tokens within the same GPU kernel execution, reallocating resources to maintain inference SLOs without the overhead of context switching or reconfiguration. During normal load, finetuning tasks opportunistically consume the idle capacity reserved for bursts, improving overall utilization. Since the base LLM parameters are shared and frozen during finetuning, memory overhead is minimized, enabling efficient multiplexing.

Overall, this paper makes the following key contributions:
\begin{itemize}[leftmargin=10pt,parsep=1pt,topsep=2pt]
\item We introduce FlexLLM, the first system to co-serve LLM inference and PEFT finetuning on shared GPUs through token-level computation fusion.
\item We develop static optimizations that reduce GPU memory requirements by up to 80\% through dependent parallelization and graph pruning.
\item We propose token-level finetuning with hybrid scheduling that maintains strict latency SLOs while maximizing GPU utilization.
\item We achieve 1.9$-$4.8$\times$ finetuning throughput improvements under heavy loads and 2.5$-$6.8$\times$ under light loads.
\end{itemize}
\section{Background and Challenges}
\label{sec:background}
\subsection{Parameter-Efficient Finetuning}
\label{subsec:peft}
Finetuning enables LLMs to adapt to specific domains and downstream tasks~\cite{devlin2018bert,radford2019language,brown2020language}. However, full finetuning introduces significant computational and memory overheads. To address this, {\em parameter-efficient finetuning} (\peft) methods~\cite{ding2023parameter} have been proposed. 
These include prompt embeddings~\cite{liu2023gpt,li2021prefix,lester2021power}, adapter modules, low-rank decomposition (LoRA)~\cite{hu2021lora}, and (IA)$^3$ \cite{liu2022few} scaling of the pre-trained weights. Recent unified approaches also combine various \peft methods using heuristics \cite{he2021towards} or neural architecture search \cite{zhou2023autopeft,zoph2016neural,mao2021unipelt}. While \peft approaches minimize trainable parameters, reducing parameter count is sometimes not enough to reduce the memory footprint.

\begin{figure*}
    \centering
    \includegraphics[width=\textwidth]{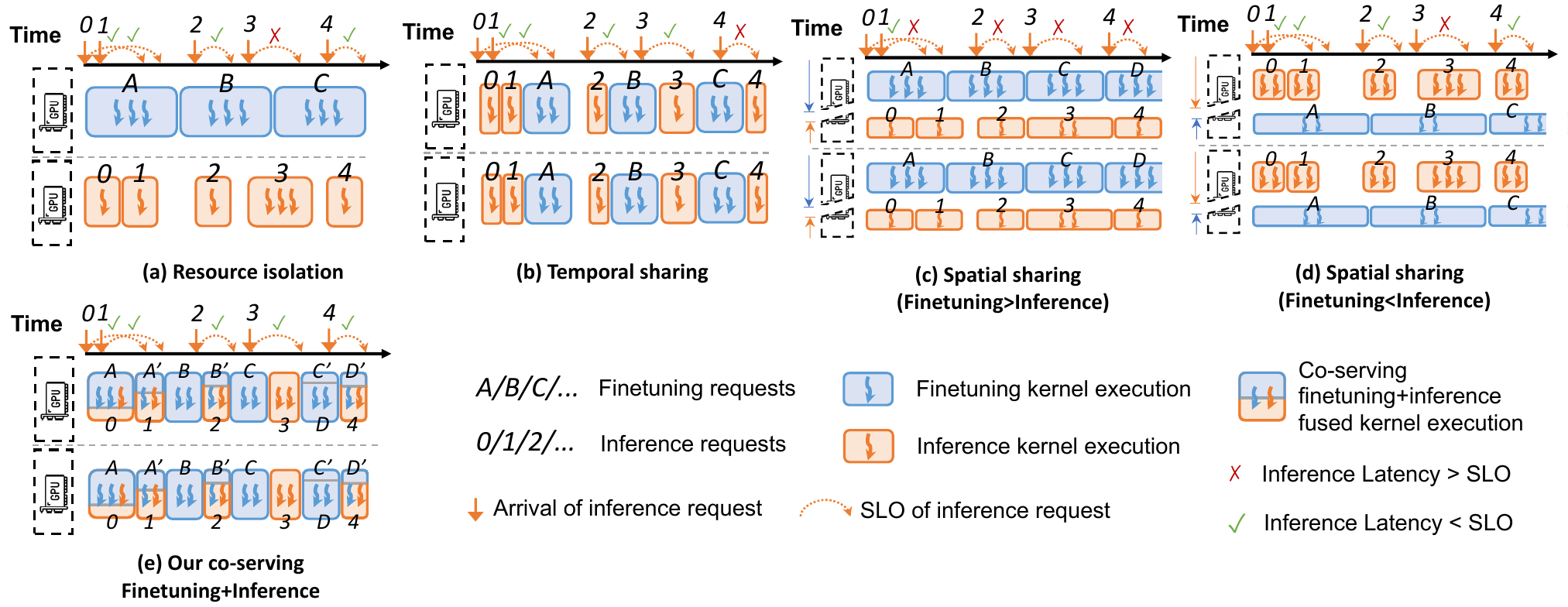}
    \caption{Comparing different resource sharing approaches for serving finetuning and inference. For spatial sharing and co-serving, the height of rounded rectangles illustrates the splitting ratio of GPU resources (e.g., streaming multi-processors).}
    \label{fig:baseline}
\end{figure*}

\subsection{\peft-based LLM Inference}
\label{subsec:llm_serving}

LLM inference is increasingly essential in research and industrial applications. Recent systems serve multiple \peft-based LLMs simultaneously. For example,PetS~\cite{zhou2022pets} decouples linear layers into shared and task-specific operations, optimizing scheduling for improved throughput.
More recent works focus on optimizations that are specific to generative decoding models. Punica~\cite{chen2023punica} introduces specialized CUDA kernels for multi-tenant LoRA serving. S-LoRA~\cite{sheng2023s} proposes unified paging and heterogeneous batching for thousands of LoRA components. dLoRA~\cite{wu2024dlora} improves efficiency through dynamic batching adjustments.

\subsection{Challenges in Co-serving Inference and Finetuning}
\label{subsec:challenges}
Co-serving inference and finetuning tasks within a single system, as envisioned by \sys, presents significant challenges due to their differing workload characteristics. \Sys must address three key challenges.

First, \emph{how can inference and finetuning tasks be co-located while minimizing the memory footprint}?  
Traditional multi-tenancy ML scheduling systems share resources across tasks and models~\cite{xiao2018gandiva}, but suffer from the high memory demands associated with LLMs.
Furthermore, existing finetuning systems usually retain all intermediate activations similar to training systems to enable gradient computation during backpropagation, which substantially increases their resource consumption compared to pure inference tasks. This discrepancy makes it challenging to manage system resources between finetuning and inference tasks.

\Sys addresses this by co-locating finetuning and inference tasks to share backbone LLM parameters and activation buffers, using {\em dependent parallelization} to optimize the placement of trainable {\em bypass networks} (e.g., LoRA adapters~\cite{hu2021lora}) in a distributed fashion (see $\S$~\ref{subsec:parallelization}) and {\em graph pruning} to eliminate unnecessary data dependencies for frozen parameters (see $\S$~\ref{subsec:pruning}). These optimizations reduce finetuning memory overhead by up to 8$\times$.

Next, \emph{how can we run inference and finetuning tasks simultaneously while minimizing the interference with inference latency and SLO attainment}? 
Previous work on traditional ML workloads~\cite{romero2021infaas} has shown that co-locating inference with training can significantly degrade inference performance.
Unlike conventional training workloads that require coarse-grained temporal or spatial resource sharing~\cite{choi2022serving}, LLM finetuning provides a unique opportunity to batch base LLM computation and parameter access across requests from different finetuned variants~\cite{wu2024dlora}. However, due to differences in execution logic---auto-regressive decoding in inference versus backpropagation in finetuning---merging the two types of tasks remains difficult.
To overcome this challenge, \Sys introduces {\em token-level finetuning}, decomposing finetuning sequences into smaller units ($\S$\ref{subsec:incremental_finetuning}) that, during the forward pass, follow the same execution order as inference tokens. In this way, both inference and finetuning tokens can be processed together using fused GPU kernels. For the backward pass, \sys launches separate GPU streams for finetuning tokens and adopts a layer-wise execution strategy to minimize memory usage.

Finally, \emph{how can we handle fluctuating inference workloads while preserving their latency SLO}?
Given the dynamic nature of inference request arrival patterns, provisioning for peak demand often results in significant resource under-utilization.
\Sys's fine-grained, token-level execution strategy enables quick adaptation to load fluctuations.
To achieve this, \sys introduces a {\em hybrid token scheduler} (Section~\ref{subsec:scheduler}) that prioritizes inference tokens for SLO compliance while opportunistically inserting finetuning tokens in a best-effort manner to maximize GPU utilization.
\section{Co-serving LLM Inference and PEFT}
\label{sec:coserving}

Previous studies have proposed different methods for scheduling heterogeneous workloads (in our case, inference and \peft) on shared GPU resources. These methods can be classified into three categories: \textit{resource isolation}, \textit{temporal sharing}, and \textit{spatial sharing}.
This section discusses the limitations inherent to these conventional approaches when applied to co-serving LLM inference and \peft. It also highlights the fundamental distinctions between \sys's co-serving solution and previous work.
\Cref{fig:baseline} depicts the GPU execution timelines associated with different scheduling approaches for managing inference and finetuning tasks.

\paragraph{Resource isolation.}
A common approach to support both inference and finetuning tasks is \textit{resource isolation}, which partitions the available GPUs into two distinct groups, each dedicated to handling one type of workload (see Figure~\ref{fig:baseline}(a)).
This approach ensures that throughput-intensive finetuning tasks do not interfere with latency-critical inference tasks. 
When finetuning an LLM, a dataset of requests is provided to train the \peft layers, with all requests submitted to \sys simultaneously, making it easier to build larger batches that can be processed sequentially. In contrast, inference requests are submitted independently by users who expect timely processing within predetermined SLOs, with the arrival times of these requests unpredictable.

Resource isolation can introduce significant GPU under-utilization, particularly in scenarios with variable request arrival rates. This is because finetuning and inference requests cannot be batched together if the number of pending requests of either type is insufficient to fully utilize a batch. With fewer GPUs available to parallelize inference computations, it may not be possible to serve requests fast enough to meet their SLO (e.g., request $r_3$ in Figure~\ref{fig:baseline}(a)).

\paragraph{Temporal sharing.}
Compared to resource isolation, \textit{temporal sharing}~\cite{xiao2018gandiva,xiao2020antman} improves GPU utilization by allowing different tasks to share GPU resources through time slices of different lengths depending on various parameters. 
This method allows both inference and finetuning requests to access all GPUs within the cluster in turns, as shown in Figure~\ref{fig:baseline}(b). By using all GPUs to parallelize computation (e.g., with data parallelism or tensor model parallelism), the latency for each request can be significantly reduced, thus aiding in meeting inference SLOs. For example, while inference request $r_3$ fails to complete in the scenario illustrated in Figure~\ref{fig:baseline}(a) due to resource limitations, it successfully executes in Figure~\ref{fig:baseline}(b) by leveraging more resources. However, the arrangement of time slices can also be harmful because of the unpredictable arrival times of inference requests, which may have a stringent latency requirement that cannot be met (e.g., $r_4$ has been delayed in Figure~\ref{fig:baseline}(b) by pre-scheduled finetuning requests).

\paragraph{Spatial sharing.}
\textit{Spatial sharing} offers an alternative by enabling simultaneous execution of different tasks at the kernel level, where each task utilizes a specific portion of the GPU resources (i.e., streaming multi-processors)
This can be implemented on NVIDIA GPUs through multi-stream programming or by utilizing the Multi-Process Service (MPS)~\cite{mps}. Furthermore, on the latest Hopper and Ampere architectures, spatial sharing benefits from the Multi-Instance GPU (MIG)~\cite{mig}, which allows for dynamic adjustments of GPU partition configurations.
However, these adjustments often lack the flexibility and efficiency required to handle dynamic inference workloads effectively~\cite{dhakal2020gslice,yu2022survey,migpartition}. Figures~\ref{fig:baseline}(c) and (d) show two scenarios where a substantial portion of GPU resources is allocated to either finetuning or inference tasks. 
While over-provisioning for inference tasks may appear to provide better SLO guarantees, it still falls short during peak demand periods such as the unexpected urgent request $r_3$. Additionally, over-provisioning wastes resources and significantly slows down the finetuning process. 
Therefore, relying solely on spatial sharing to manage finetuning and inference tasks often results in suboptimal performance ~\cite{choi2022serving,zhao2023muxflow}.

\paragraph{\sys's co-serving approach.}
As shown in Figure~\ref{fig:baseline}(e), \sys employs a co-serving approach that uses a fine-grained scheduling mechanism to adaptively manage both inference and finetuning requests.

Unlike temporal or spatial sharing approaches, which schedule GPU kernels of existing systems independently, \sys integrates inference and finetuning kernels. This integration reduces the overhead associated with kernel launches and minimizes accesses to GPU device memory to retrive model weights.
In each iteration, \sys dynamically adjusts the allocation of inference and finetuning tokens, strategically balancing the need to meet the SLOs of inference requests with the goal of maximizing GPU utilization. This adaptive approach ensures that \sys adheres to the requirements of inference requests while optimizing the use of available GPU resources.

\section{System Overview}
\label{sec:overview}

\begin{figure}[t]
    \centering
    \includegraphics[width=\linewidth]{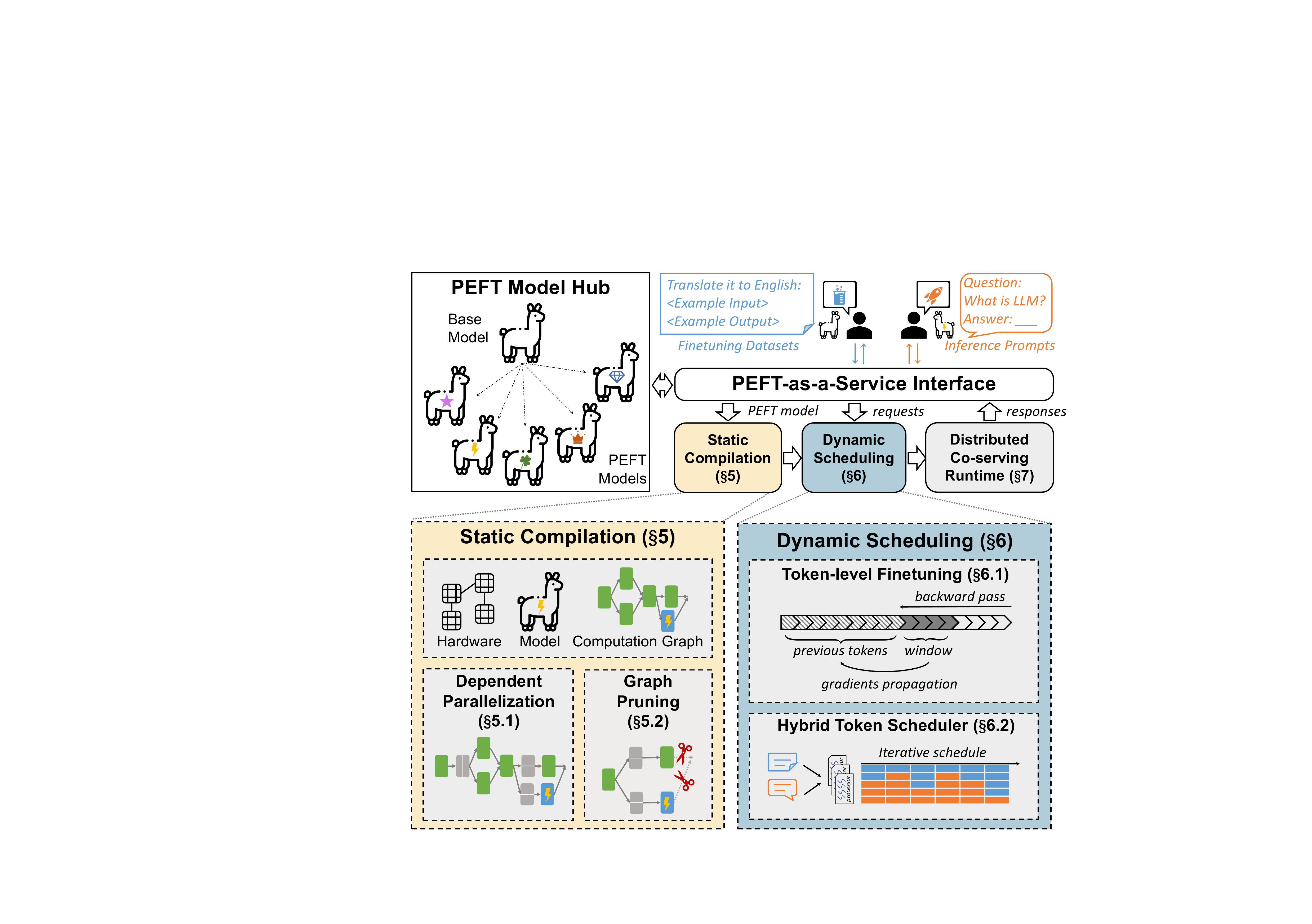}
    \caption{An overview of \sys.}
    \label{fig:overview}
\end{figure}

Figure~\ref{fig:overview} shows an overview of \sys, which co-serves both \peft and inference requests for LLMs. Users can choose to use either finetuning or inference services on a model from the PEFT {\em model hub}, which stores the backbone LLM and all finetuned variants.
The programming interface for both inference and finetuning requests is unified through a \textit{PEFT-as-a-Service (PaaS) interface}.
\sys comprises three main components.
First, for finetuning requests that execute in parallel across multiple GPUs, the static compilation module (\S\ref{sec:compiler}) generates a parallel computation graph. This graph specifies the execution of the PEFT model over the distributed environment and optimizes the graph by pruning unnecessary tensors for memory saving. 
Second, the dynamic scheduling module (\S\ref{sec:scheduling}) adapts a token-level finetuning mechanism. It mixes inference and finetuning tokens with a hybrid scheduling policy.
Finally, the computation graph and schedule plan are executed by \sys's distributed co-serving runtime (\S\ref{sec:impl}).

\begin{figure}[t]
    \centering
    \includegraphics[width=\linewidth]{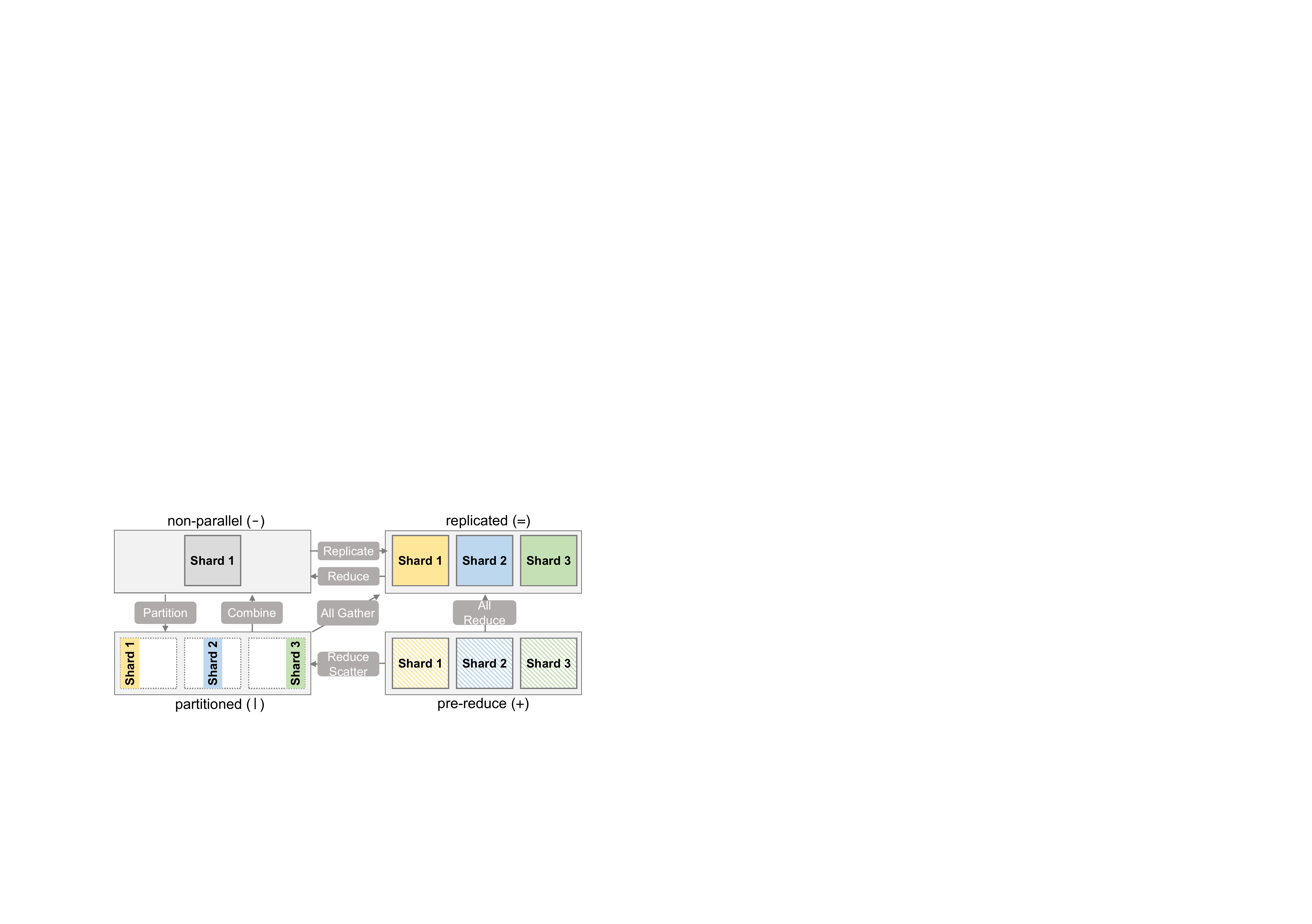}
    \caption{Four possible parallel states for a tensor dimension and their transitions. For each parallel state, the symbol in parenthesis shows the notation \sys used to represent it.}
    \label{fig:pcg_states}
\end{figure}

\begin{figure*}[t]
    \centering
    \includegraphics[width=\textwidth]{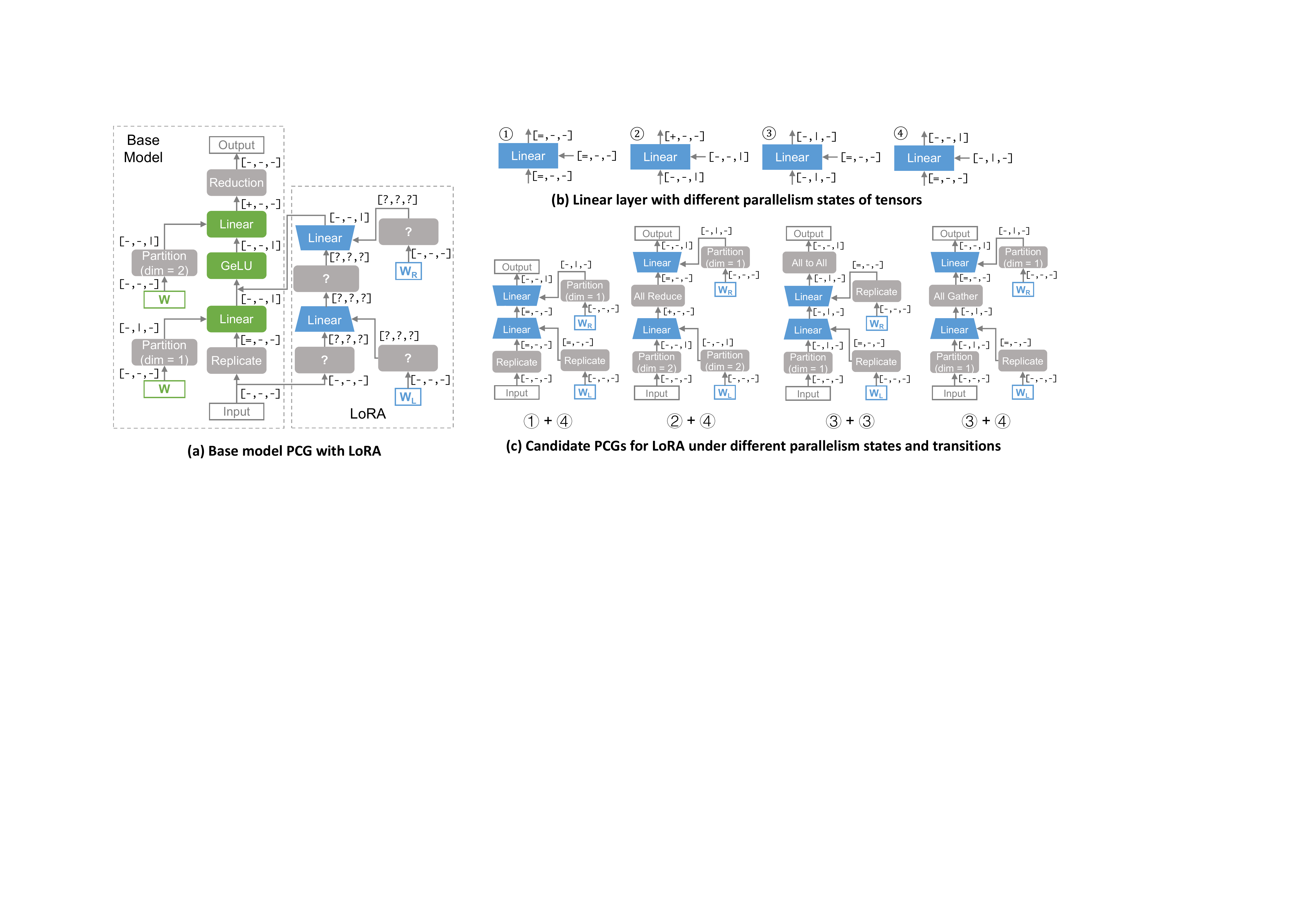}
    \caption{Illustration of \sys's different dependent parallelization strategies with the LoRA example. Each green (or gray) box indicates a compute (or parallelization) operator, and each edge between operators represents a parallel tensor, and the parallelization states of the tensor's dimensions are shown next to the edge.
    }
    \label{fig:candidate_pcg}
\end{figure*}

\subsection{\peft-as-a-Service Interface}
\label{subsec:paas_apis}

To facilitate the sharing of the backbone LLM across different tasks, \sys represents a \peft model as a sequence of {\em bypass networks} attached to the backbone LLM.
Each bypass network takes a {\em single} tensor from the backbone LLM as input and produces a {\em single} output tensor, which is added to one tensor of the backbone LLM.
Let $X$ and $Y$ denote the input and output tensor of the bypass network, and let $f_B(X)$ and $f_A(X)$ denote the neural architecture of the backbone and bypass network for calculating $Y$. The bypass network can be formulated as $Y = f_B(X) + f_A(X)$.
All existing \peft methods can be represented in this format. For instance, the (IA)$^3$ architecture modifies the topology of the backbone LLM by inserting an elementwise multiplication operator. It can be transformed into \sys's bypass network format since $Y = X\odot W = X + X \odot (W - O)$, where $\odot$ is elementwise multiplication and $O$ is a matrix of the same shape as $W$ whose elements are all one.
A significant benefit of this interface is that all \peft models preserve the neural architecture of the backbone LLM, which enables \sys to fuse the computation graphs of different \peft models.
\sys provides a unified user interface to launch inference and finetuning tasks by specifying a \peft model and committing different types of input data, such as finetuning datasets or inference prompts.

\section{Static Compilation}
\label{sec:compiler}

Once a \peft model is registered, \sys compiles it into a parallel computation graph (PCG)~\cite{unger2022unity}. To optimize finetuning performance, \sys applies two key static compilation optimizations to discover an optimized PCG: {\em dependent parallelization} (\S\ref{subsec:parallelization}) and {\em graph pruning} (\S\ref{subsec:pruning}).

\subsection{Dependent Parallelization}
\label{subsec:parallelization}

To minimize memory overhead, \sys allows finetuning and inference tasks to utilize the same backbone LLM and share memory allocated for both model weights and intermediate activations. This approach avoids redundant memory consumption while requiring the parallelization strategies of the \peft models to be compatible with that of the backbone LLM, a task termed {\em dependent parallelization}.

Existing auto-parallelization approaches like Alpa~\cite{zheng22-alpa} and Unity~\cite{unger2022unity} can identify optimal parallelization strategies for backbone LLMs.
\Sys extends the optimization target to also include the bypass networks in \peft models.
To achieve this goal, \sys generalizes the {\em parallel computation graph} (PCG) abstraction introduced in Unity~\cite{unger2022unity} by associating a {\em state} field with tensor dimension.

\Cref{fig:pcg_states} shows the four possible tensor states when parallelizing a tensor, including {\em non-parallel} ({\tt -}), {\em partitioned} ({\tt |}), {\em replicated} ({\tt =}), and {\em pre-reduce} ({\tt +}), as well as the state transitions.
Given the fixed parallelization of the backbone LLM, \sys searches for an optimal PCG for each bypass network by enumerating possible parallelization operators---{\tt partition}, {\tt combine}, {\tt replicate}, and {\tt reduce}---inferring output tensor states, and validating their compatibility.
\Cref{fig:candidate_pcg} shows four candidate PCGs discovered by \sys for parallelizing a LoRA network. To select the best strategy, \sys reuses Unity's profiling-based cost model~\cite{unger2022unity} and chooses the candidate PCG with the lowest estimated execution cost.

\begin{algorithm}[t]
\caption{Static graph pruning. For an operator $n$, $\textproc{UpdateInput}(n, \m{O}(n))$ returns a set of input tensors needed in order only to compute $\m{O}(n)$ of the operator.}
\label{alg:graph_pruning}
\begin{algorithmic}[1]
\State {\bf Inputs:} PCG of a PEFT model $\m{G}$
\State {\bf Outputs:} $\m{A}$ is a set of tensors to be memorized, $\m{R}$ is a set of tensors to be rematerialized
% \State \Comment{Step 1: computation graph pruning}
\LeftComment{Step 1: computation graph pruning}
\State $\overline{\m{G}} = \Call{ReverseAutoDiff}{\m{G}}$
\State $\m{Q} = \emptyset$ \Comment{$\m{Q}$ is a queue of updated operators}
\For{operator $n \in \overline{\m{G}}$}
%\State $\m{O}(n) = \m{O}(n) \setminus \m{B}$ \Comment{$\m{B}$ is the set of gradients to the backbone LLM}
\For{output tensor $t \in \m{O}(n)$}
\If{$t$ is the weight gradient of the base LLM}
\State $\m{O}(n) = \m{O}(n) \setminus \{t\}$
\State $\m{I}(n) = \Call{UpdateInput}{n, \m{O}(n)}$
\State $\m{Q}$.push\_back($n$)
\EndIf
\EndFor
\EndFor
\While{$\m{Q}$ is not empty}
\State $n = \m{Q}$.pop\_front()
\For{output tensor $t \in \m{O}(n)$}
\If{$\nexists u. t \in \m{I}(u)$ }
\State $\m{O}(n) = \m{O}(n) \setminus \{t\}$
\State $\m{I}(n) = \Call{UpdateInput}{n, \m{O}(n)}$
\State $\m{Q}$.push\_back($n$)
\EndIf
\EndFor
\EndWhile
\State $\m{A} = \emptyset$
\For{operator $n \in \m{G}$}
\For{tensor $t \ in \m{O}(n)$}
\If{$\exists u \in \overline{\m{G}}. t \in \m{I}(u)$}
\State $\m{A} = \m{A} \cup \{ t\}$
\EndIf
\EndFor
\EndFor
% \Statex \Comment{Step 2: opportunistically rematerializing tensors}
\LeftComment{Step 2: opportunistically rematerializing tensors}
\For{tensor $t \in \m{A}$}
\State Let $n$ be the operator that outputs $t$ (i.e., $t \in \m{O}(n)$)
\If{$\m{I}(n) \subseteq \m{A}$ and $\Call{Cost}{n} < threshold$}
\State $\m{A} = \m{A} \setminus \{t\}$, $\m{R} = \m{R} \cup \{t\}$
\EndIf
\EndFor
%\State \Comment{Step 3: tensor compression}
\State {\bf return} $\m{A}, \m{R}$
%\Function{aaa}{}
%\EndFunction
\end{algorithmic}
\end{algorithm}

\begin{figure}[t]
    \centering
    \includegraphics[width=0.8\linewidth]{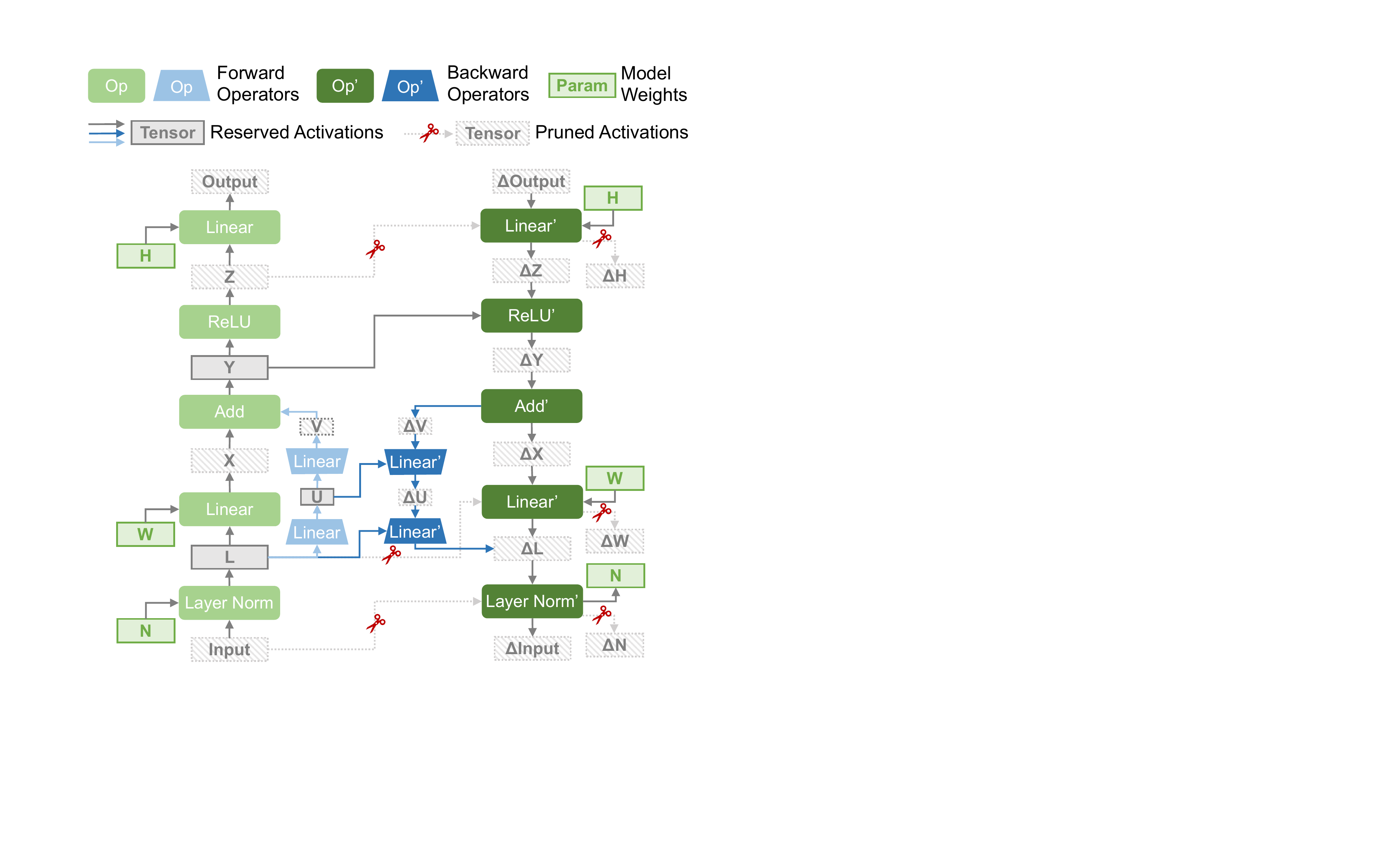}
    \caption{Static graph pruning for an MLP model with LoRA.}
    \label{fig:graph_pruning}
\end{figure}

\begin{figure*}[t]
    \centering
    \includegraphics[width=0.85\textwidth]{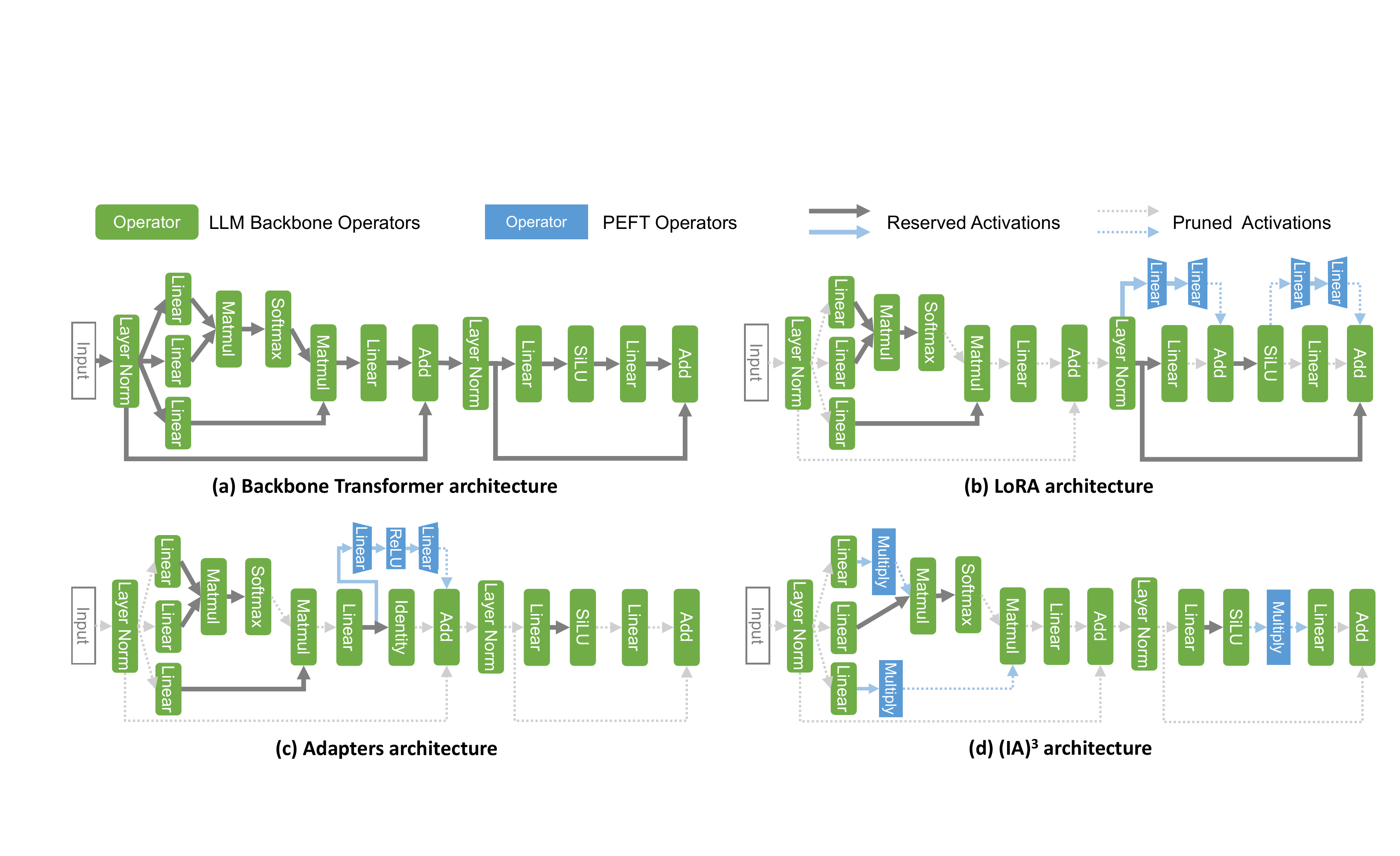}
    \caption{An overview of existing parameter-efficient finetuning (PEFT) methods. Green boxes show the operators of the backbone LLM, while blue boxes are the operators introduced by different PEFT methods. Arrows represent intermediate activations to be reserved and the dashed arrows demonstrate they are pruned by \sys.}
    \label{fig:peft-methods}
\end{figure*}

\subsection{Graph Pruning}
\label{subsec:pruning}

\begin{figure*}[t]
    \centering
    \includegraphics[width=\textwidth]{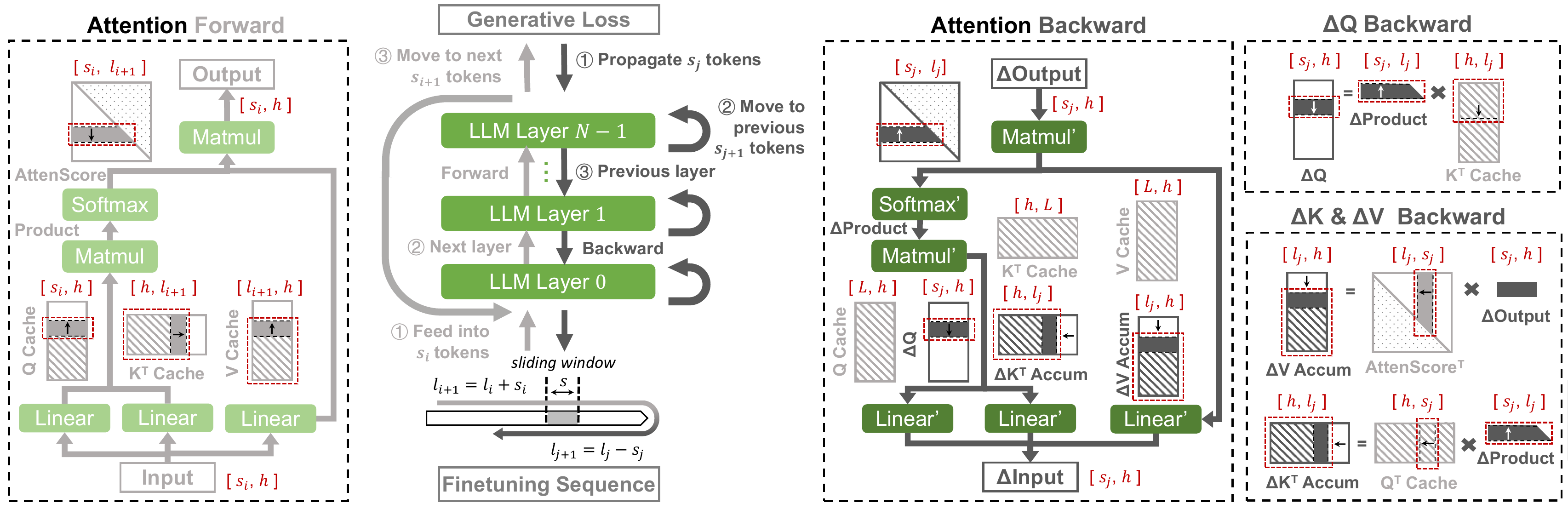}
    \caption{Illustration of attention module's forward and backward execution with \sys's token-level finetuning mechanism.}
    \label{fig:token-level-finetuning}
\end{figure*}

\sys's {\em graph pruning} algorithm takes as input a PEFT model and its backbone LLM, and outputs a minimal set of intermediate activations that must be reserved to perform finetuning of the bypass networks.
This pruning process requires reasoning over data dependencies between operators, specifically their input and output tensors. Importantly, this pruning process preserves model quality since gradients with respect to frozen parameters are mathematically unnecessary for PEFT optimization and do not affect the final trained model. To facilitate the analysis of data dependencies, \sys introduces additional notation in the PCG representation.
Let $\m{G}=(\m{N}, \m{E})$ denote the PCG of a PEFT model, where each node $n \in \m{N}$ represents a tensor algebra (or parallelization) operator, and each edge $e=(n_1, n_2) \in \m{E}$ is a tensor shared between operators. For a node $n$, let $\m{I}(n)$ and $\m{O}(n)$ denote its set of input and output tensors.
By definition, $(n_1, n_2) \in \m{E}$ if and only if $\m{O}(n_1) \cap \m{I}(n_2) \neq \emptyset$, i.e., when at least one output tensor of $n_1$ serves as an input to $n_2$.

The design of \sys's graph pruning algorithm is based on two key observations.
First, conventional ML training procedures maintain all intermediate activations during the forward pass to compute gradients in the backward pass.
However, due to the linear algebra nature of most operators, these intermediate activations are primarily used to compute gradients with respect to the base LLM's trainable parameters. In the case of \peft, the backbone LLM parameters are frozen, and only the parameters in the bypass networks are updated.
This insight allows \sys to eliminate most intermediate activations, retaining only those necessary for computing gradients of the bypass networks. 
Second, different PEFT methods connect bypass networks at varying locations within the backbone LLMs, and therefore require different subsets of intermediate activations for backpropagation.
\Cref{fig:peft-methods} illustrates examples of how various \peft methods attach their bypass networks to the base LLM.

Building on these observations, \sys performs {\em static} graph pruning during the construction of the PCG for each registered \peft model, and {\em dynamical} scheduling to manage finetuning and inference tasks, along with corresponding memory allocations (see \Cref{sec:scheduling}).
\Cref{alg:graph_pruning} shows \sys's graph pruning algorithm, with \Cref{fig:graph_pruning} illustrating the process for an MLP model with LoRA. Given a PEFT model's PCG $\m{G}$, the algorithm generates an execution plan specifying which intermediate activations to cache for efficient backpropagation.

The algorithm constructs the backward graph $\overline{\m{G}}$ using reverse-mode automatic differentiation~\cite{hogan2014fast}, then prunes gradients for frozen base LLM weights. It iteratively processes operators to discover pruning opportunities, ultimately identifying the minimal set of activations $\m{A}$ required for the backward pass. \sys combines this with {\em rematerialization} and {\em activation compression}.

\paragraph{Rematerialization and compression.}
\sys applies tensor rematerialization~\cite{checkmate}, selectively discarding tensors in the forward pass and recomputing them during backpropagation. For each tensor $t \in \m{A}$ after graph pruning, \sys rematerializes $t$ if all input tensors are stored and recomputation has low overhead. Additionally, \sys opportunistically applies lossless compression when operators like {\tt ReLU} don't require access to original input tensors. 
$y=\texttt{ReLU}(x) = \max{(x, 0)}$ whose derivative is $\partial y/\partial x = 1$ for $x>0$ and $0$ for $x\leq0$. Therefore, instead of storing the original input tensor $x$, \sys keeps the bitmask of $x$.

\section{Dynamic Scheduling}
\label{sec:scheduling}
Once static compilation produces an executable generalized PCG for a \peft model, \sys must determine an execution schedule capable of concurrently handling both finetuning and inference requests.
This is particularly challenging due to the distinct characteristics of these two workloads. 
First, inference requests arrive in an online and highly dynamic manner, requiring the system to rapidly adapt GPU resource allocation to maintain low-latency responses. In constrast, finetuning involves both a forward pass---similar to inference---and a backward pass to update model parameters. 
Additionally, auto-regressive inference for generative LLMs proceeds in a token-by-token fashion, while finetuning is typically performed at the sequence level to maximize throughput.
To address these challenges, \sys introduces a novel \textit{token-level finetuning} mechanism along with a \textit{hybrid token scheduler}.

\begin{figure}[t]
    \centering
    \includegraphics[width=0.8\linewidth]{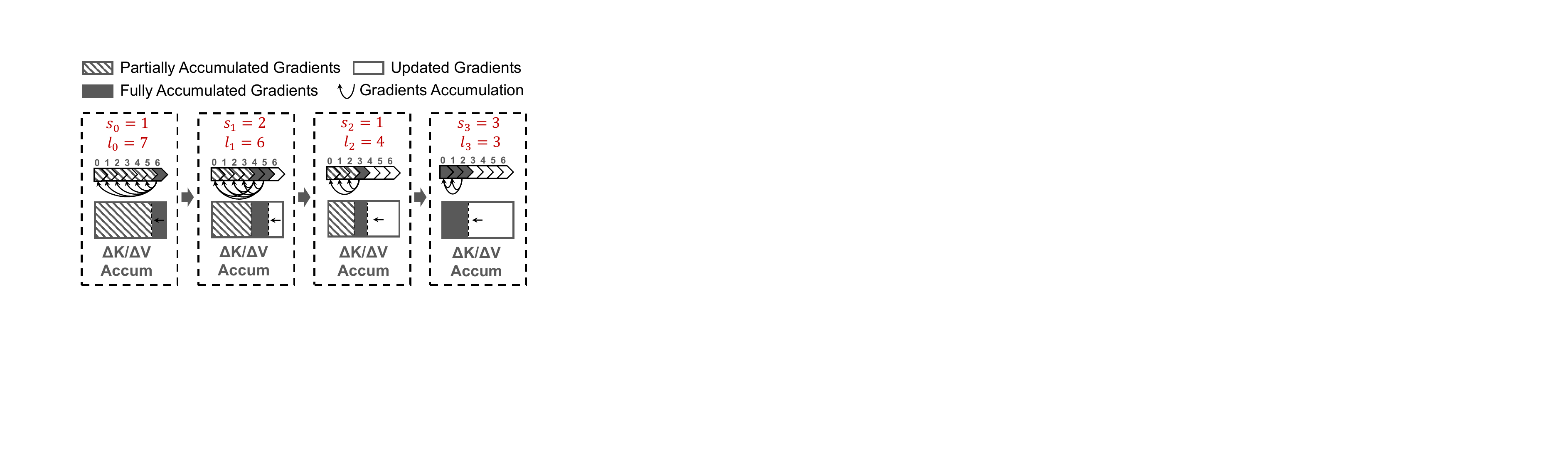}
    \caption{Illustration of KV gradients accumulation in the backward pass of \sys's token-level finetuning mechanism, where $s_j$ and $l_j$ are the window length and the starting position of the $j$-th slice of the sequence.}
    \label{fig:token-level-backword}
\end{figure}

\subsection{Token-Level Finetuning}
\label{subsec:incremental_finetuning}

Processing a finetuning request typically involves performing both forward and backward passes across all model layers for the entire input sequence.
When co-served with inference requests, this sequence-level execution can significantly interfere with the token-wise inference kernels and degrade system goodput---this is, the effective inference throughput that meets latency SLOs.

While sequence-level parallelism~\cite{megatron-sp, deepspeed-ulysses} could help reduce the added finetuning latency per GPU, this approach would suffer from coarse-grained scheduling, forcing the system to wait for entire finetuning sequences to complete before reallocating resources to handle inference bursts. 

To address this problem and maximize GPU utilization, \sys introduces a \textit{token-level finetuning} mechanism, detailed in \Cref{alg:token-level-finetuning}. The key idea is to decompose the finetuning computation into smaller steps using a dynamic sliding window over tokens. This window spans both the forward and backward passes, with its size dynamically determined by the hybrid token scheduler (line 4 and 15) to ensure that inference requests continue to meet their SLO constraints (\S\ref{subsec:scheduler}). 

During the forward pass (lines 3-11), the finetuning sequence is partitioned into windows of $s_i$ tokens (line 5), which are incrementally fed into the model. The model processes each token window layer by layer (lines 7-8), computes the generative loss for the window (line 10), and then advances to the next set of tokens (line 11).
To preserve the semantic of full sequence-level finetuning, \sys caches key and value tensors---similar to incremental decoding in inference~\cite{yu2022orca}---as well as query tensors, which are reused during backward attention computations. This caching follows the {\em causal mask} constraint inherent in LLMs.

During the backward pass (lines 12-21), \sys executes the model layers in reverse order (line 13).
Within each layer, the finetuning request is again divided into token windows (lines 14-16). 
However, unlike the relatively straightforward forward pass, token-level backward execution of the attention module is more complex due to auto-regressive token dependencies.

Figure~\ref{fig:token-level-finetuning} illustrates the forward and backward PCGs of the attention module. When computing gradients for $s_j$ tokens, the resulting gradients $\Delta Q$, $\Delta K$, $\Delta V$ have shapes $[s_j, h]$ for queries but $[l_j, h]$ for keys and values due to autoregressive attention dependencies. \sys accumulates KV gradients across steps, applying them only after complete backward traversal. This accumulation strategy minimally increases memory consumption due to layer-wise execution enabling workspace memory reuse.

\begin{algorithm}[t]
\caption{Token-level finetuning mechanism.}
\label{alg:token-level-finetuning}
\begin{algorithmic}[1]
\State {\bf Inputs}: PEFT model $M$ depth of $N$; Finetuning request $r$ with sequence length of $L$
\State {\bf Outputs}: PEFT model gradients $\Delta M$
% \State $i\gets 0, l_0 \gets 0$ \Comment{Forward pass}
\For{$i\gets 0, l_0 \gets 0$; $l_i<L$; $i\texttt{++}$} \Comment{Forward pass}
\State $s_i\gets \Call{HybridTokenScheduler}{l_i, L}$ 
\State $r_i \gets \Call{Slice}{r, l_i, s_i}$
\State $X_{0,i} \gets \Call{Tokenize}{r_i}$
\For{model layer index $n \in \Call{Range}{0, N}$}
\State $X_{n+1, i}, Q_{n, i}, K_{n, i}, V_{n, i} \gets \Call{Forward}{M_n, X_{n, i}}$
\State $QKVCache_n\gets \Call{Append}{Q_{n, i}, K_{n, i}, V_{n, i}}$
\EndFor
\State $Loss_{i} \gets \Call{GenerativeLoss}{X_{N, i}, r_i}$
\State $l_{i+1}\gets l_i + s_i$
% \State $l_{i+1}\gets l_i + s_i$, $i\gets i+1$
% \EndWhile
\EndFor
% \State $j\gets 0, l_0 \gets L$ \Comment{Backward pass}
% \While{$l_j>0$}
\State $Y_{N}\gets Loss$
\For{model layer index $n \in \Call{Range}{N-1, -1, -1}$}
\For{$j\gets 0, l_0 \gets L$; $l_j>0$; $j\texttt{++}$} \Comment{Backward pass}
\State $s_j\gets \Call{HybridTokenScheduler}{l_j, 0}$
\State $Y_{n+1, j} \gets \Call{Slice}{Y_{n+1}, l_j, s_j}$
% \State $Y_{j,n}, \Delta K_i, \Delta V_i \gets \Call{Backward}{M_n, Y_{j,n-1}, QKVCache, \Delta KVAccum}$
\State $Y_{n, j}, \Delta K_{n, j}, \Delta V_{n, j} \gets \textsc{Backward}(M_n, Y_{n+1, j},$
\Statex \hfill $QKVCache_n, \Delta KVAccum_n)$
\State $\Delta KVAccum_n\gets\Call{Add}{\Delta K_{n, j}, \Delta V_{n, j}}$
\State $G_{n, j}\gets \Call{Slice}{\Delta KVAccum_n, l_j, s_j}$
\State $\Delta M_{n} \gets \Delta M_{n}+\Call{CalculateGrads}{M_n, G_{n, j}}$
\EndFor
\State $l_{j+1}\gets l_j-s_j$
\EndFor
% \State $l_{j+1}\gets l_j-s_j, j\gets j+1$
% \EndWhile
% \Function{Build}{$\m{G}, \m{I}$}
% \EndFunction
\State {\bf return} $\Delta M$
\end{algorithmic}
\end{algorithm}

\begin{figure}[t]
    \centering
    \includegraphics[width=\linewidth]{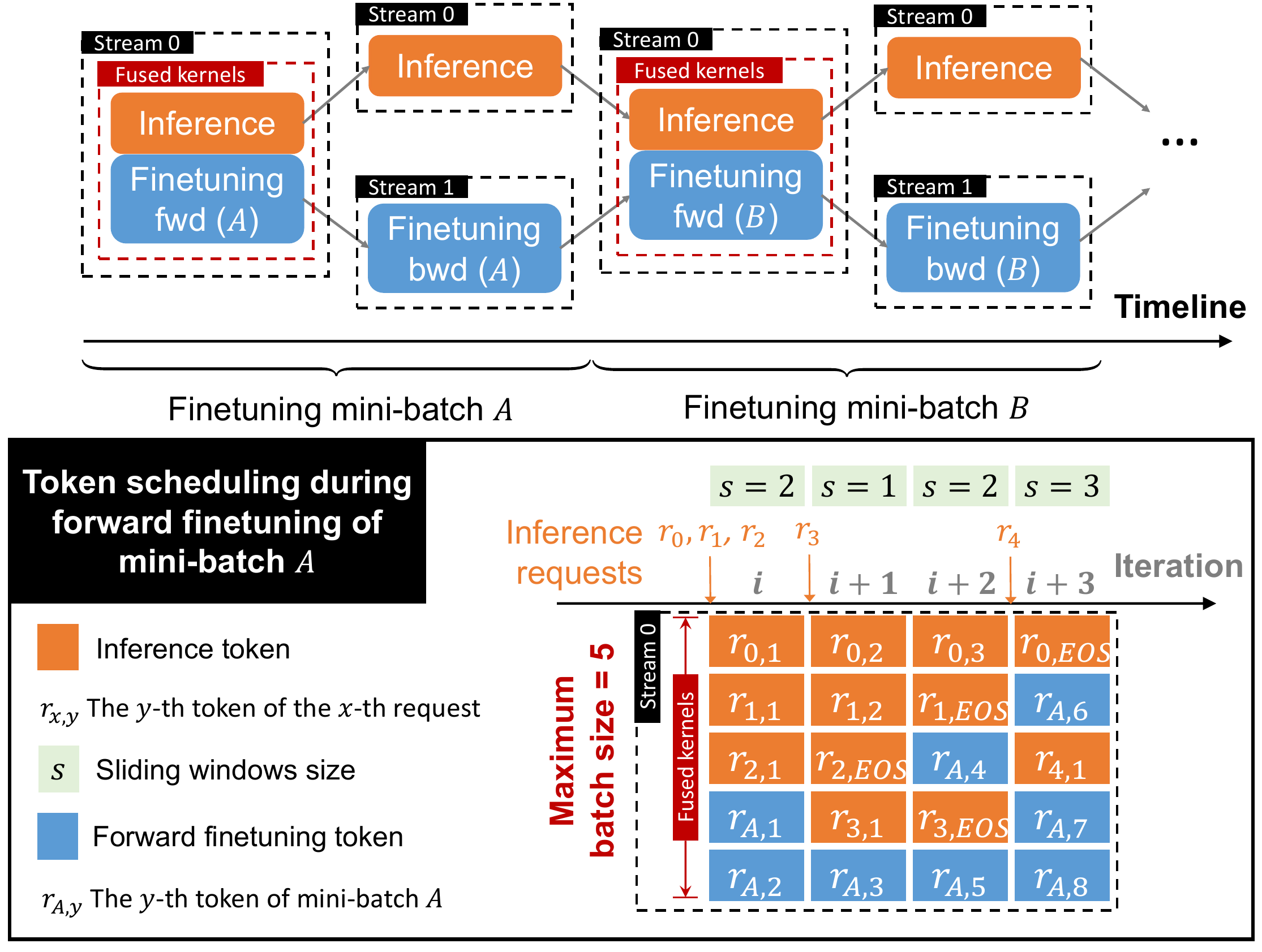}
    \caption{Execution timeline and token scheduling when co-serving inference requests and a finetuning mini-batch.}
    % Numbers in parenthesis indicate the iteration numbers for serving and PEFT operators, respectively, and arrows show data dependencies between operators. 
    \label{fig:execution_timeline}
\end{figure}

\subsection{Hybrid Token Scheduler}
\label{subsec:scheduler}

\Sys's \textit{hybrid token scheduler} coordinates token-level GPU execution in two stages. 
First, \Sys determines the scheduling of available inference requests based on specific scheduling policies. By default, \sys adopts Orca's \textit{iteration-level scheduling}~\cite{yu2022orca}, which maintains a fixed maximum batch size and dynamically replaces each completed request with a new one whenever available.
To further mitigate blocking caused by long input sequences, \sys incorporates the chunked-prefill optimization~\cite{agrawal2024taming}.

Second, after determining the inference schedule for a given iteration, \sys opportunistically appends as many finetuning tokens as possible---determined by the sliding window size $s$---to maximize GPU utilization. The number of finetuning tokens added is determined automatically using the formula $s=\argmax f(c, s)\leq$SLO, where $f(\cdot, \cdot)$ is the latency estimation function and $c$ is the number of inference tokens scheduled in the current iteration. Here $f(\cdot, \cdot)$ is derived via offline profiling of the LLM's execution~\cite{narayanan2023cheaply}.
Such window-based, best-effort scheduling mechanism ensures that \sys can automatically adapt to inference workload fluctuations, maintaining inference SLOs while opportunistically allocating compute to finetuning tasks (\S\ref{sec:exp:dynamic}).

Figure~\ref{fig:execution_timeline} illustrates the execution timeline of \Sys and shows an example a token scheduling plan for a finetuning mini-batch $A$, under a representative inference requests arrival pattern (i.e., $r_0$ to $r_4$).
To preserve the semantics of finetuning, \sys enforces the execution dependencies between the forward and backward passes of each mini-batch.
During the forward pass, \sys leverages fused GPU kernels to jointly process both inference and finetuning tokens. This approach avoids additional kernel launch overhead and is enabled by the shared token-wise computation logic between inference and finetuning. In particular, finetuning tokens follow the same execution pattern and causal masking rules as inference tokens during the prefill phase, guaranteeing compatibility at the kernel level.
For the backward pass, \sys adapts a space-sharing approach, using two separate GPU streams: one for inference tokens and another for finetuning tokens. This design allows concurrent execution without interference.

\begin{figure*}[t]
    \centering
    \includegraphics[width=0.75\textwidth]{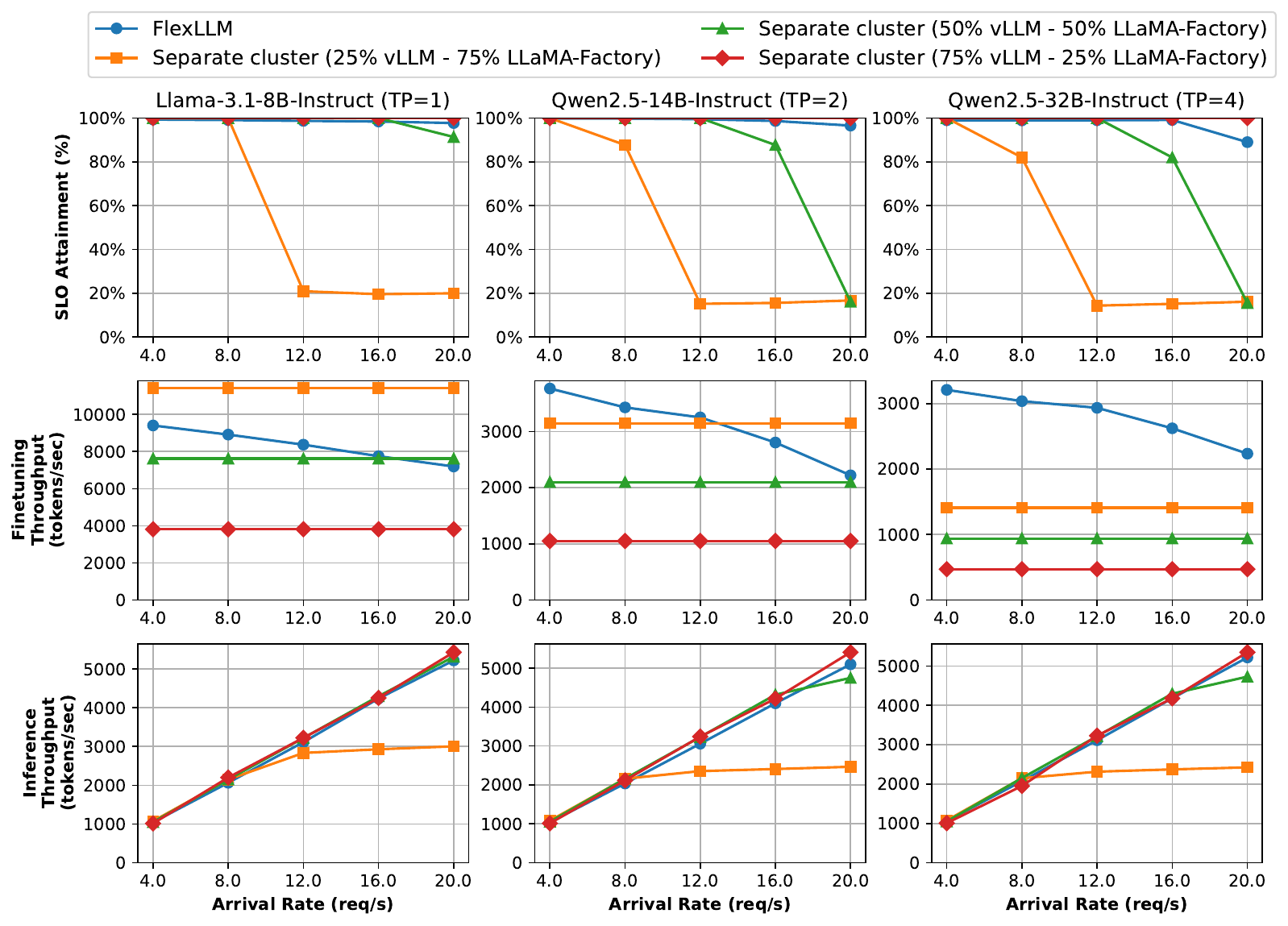}
    \caption{End-to-end comparison between co-serving and using separate resources on three models.}
    \label{fig:end-to-end}
\end{figure*}

\begin{figure*}[t]
    \centering
    \includegraphics[width=0.75\textwidth]{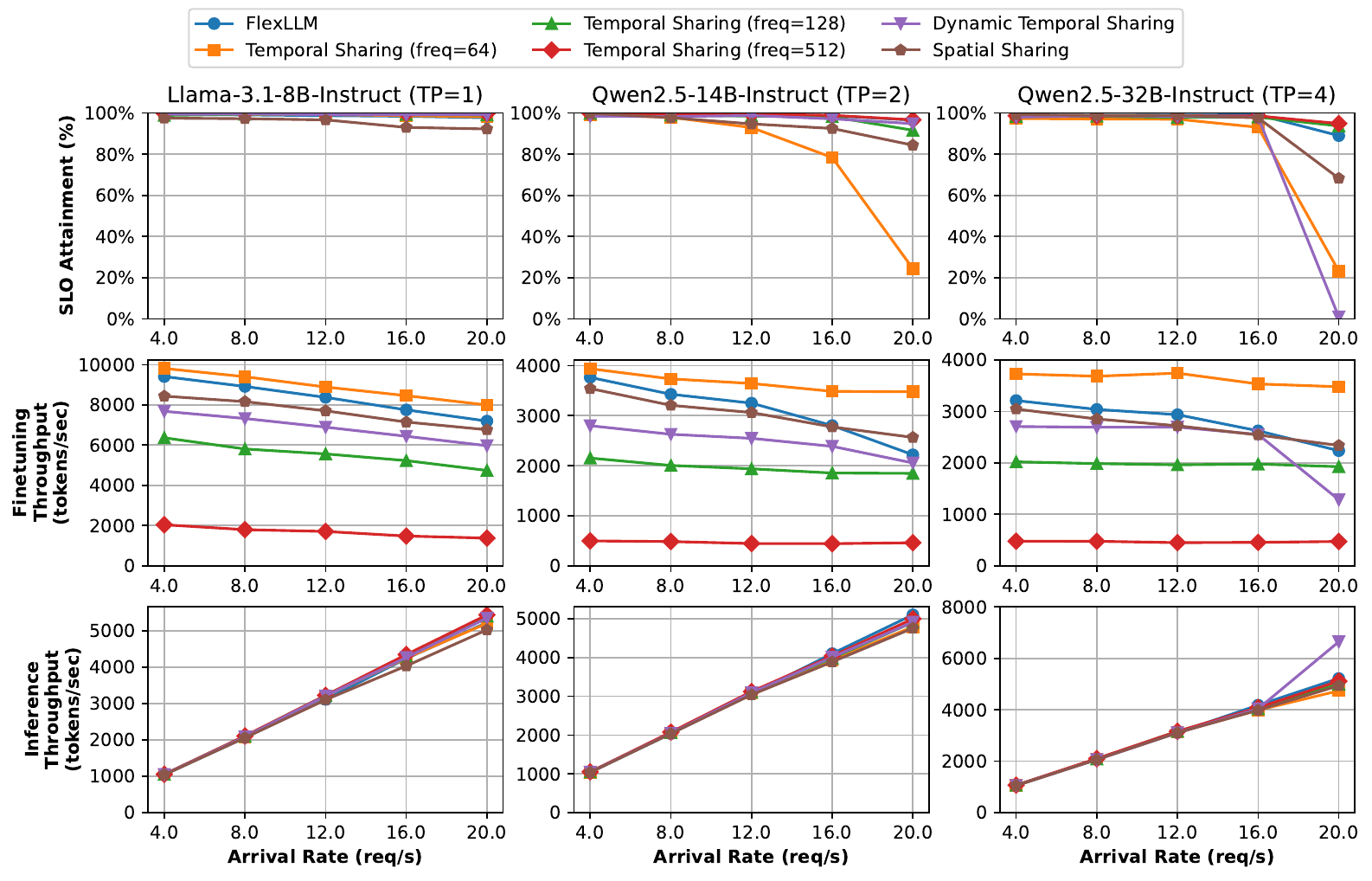}
    \caption{End-to-end comparison of co-serving with temporal and spatial sharing strategies.}
    \label{fig:temporal-spatial-sharing}
\end{figure*}

\section{Implementation}
\label{sec:impl}
\sys is built on FlexFlow Serve~\cite{miao2023specinfer} with 9K lines of C++ and CUDA code. It provides C++ and Python bindings, with full HuggingFace compatibility for \texttt{PEFT}~\cite{peft} and \texttt{transformers}~\cite{transformers} libraries. \Sys supports PEFT models with LLAMA~\cite{touvron2023llama}, GPT~\cite{brown2020language}, OPT~\cite{zhang2022opt}, Falcon~\cite{falcon40b}, or MPT~\cite{MosaicML2023Introducing} backbones, among others. 

\paragraph{Memory management.}
\Sys manages the required GPU memory using a combination of static and dynamic allocation. Static allocation reserves GPU space for the backbone weights and the KV cache required for incremental decoding during inference. Dynamic allocation is used for gradients, activations, and optimizer states. When a finetuning request is received, \Sys allocates memory during the first forward pass and reuses it during subsequent backward passes, freeing up space when it's no longer needed.
For inference memory management, FlexLLM employs paged attention~\cite{10.1145/3600006.3613165} with chunked prefill~\cite{agrawal2023sarathi} to dynamically allocate KV cache pages and minimize evictions. New inference requests are only admitted if the entire prompt can fit within available KV cache pages, preventing memory fragmentation and ensuring stable inference performance during co-serving.

\paragraph{Key-value gradient accumulator.}
To support token-level finetuning, it is necessary to accumulate key-value gradients for every preceding token in a given sequence during backpropagation. \sys utilizes static allocation to reserve space for key-value gradient accumulation. During backpropagation, the gradients for keys and values of a specific segment are obtained, and the position of this segment in the overall sequence is determined. \sys then accumulates the new key-value gradients of all subsequently scheduled segments. When a segment is scheduled during backpropagation, it includes accumulated gradients from future tokens.

\section{Evaluation}
\label{sec:eval} 
\paragraph{LLMs and PEFT models.} We evaluated \sys across three publicly available LLMs: \llama-3.1-8B~\cite{llama2}, Qwen-2.5-14B~\cite{qwen2025qwen25technicalreport}, and Qwen-2.5-32B~\cite{qwen2025qwen25technicalreport} from HuggingFace~\cite{huggingfacemodels}.
Since existing LLM serving systems only support base models and LoRA-based \peft, we focus on these configurations for fair comparison.
We applied LoRA with rank 16 to MLP down projection layers, yielding 9.4M, 14.5M, and 25.16M trainable parameters respectively.

\paragraph{Platform.} Experiments were performed on Perlmutter~\cite{perlmutter} nodes with AMD EPYC 7763 processors, \SI{256}{GB} DRAM, and four NVIDIA A100-SXM4-80GB GPUs connected via HPE Slingshot (\SI{200}{Gb/s}). We used tensor-model parallelism for LLM serving and finetuning.

\paragraph{Workload.} 
We synthesized inference workloads using ShareGPT~\cite{sharegpt} prompt/generation lengths and Azure ChatGPT production traces~\cite{burstgpt} for realistic arrival patterns, following recent work~\cite{10.1145/3600006.3613165, zhong2024distserve, agrawal2023sarathi, nanoflow, pipellm, medusa}. The original trace spans multiple days and captures real production arrival patterns from ChatGPT serving on Azure OpenAI, providing realistic inference workload characteristics including bursty request arrivals and varying concurrency patterns.  We sampled 20-minute intervals, adjusting the arrival times to simulate different average arrival rates.
We set TPOT SLOs to 50ms (8B model) and 75ms (14B/32B models) following the methodology from prior work~\cite{agrawal2023sarathi}, with 5s maximum TTFT to prevent excessive queueing.

We sampled each finetuning request from the \texttt{Sky-T1\_data\_17k} dataset~\cite{novasky2025dataset}, used to finetune the \texttt{Sky-T1-32B-Preview} reasoning model~\cite{novasky2025model}, a finetuned Qwen2.5 variant with performance comparable to OpenAI's o1-preview~\cite{o1-preview} on math and coding benchmarks. We truncated sequences to 8192 tokens and used the Adam optimizer~\cite{kingma2014adam}.

\subsection{End-to-end Comparison}

We compare \sys's co-serving approach with existing approaches that employ separate clusters for inference and finetuning tasks. In the separate cluster approach, we use vLLM~\cite{10.1145/3600006.3613165} and LlamaFactory~\cite{llamafactory} as the inference and finetuning systems, respectively, as they represent state-of-the-art solutions for these workloads. We selected vLLM over alternatives like DeepSpeed-MII~\cite{deepspeed-mii} and TensorRT-LLM~\cite{tensorrt-llm} because recent vLLM optimizations have closed previous performance gaps, while PipeSwitch~\cite{bai2020pipeswitch} lacks support for autoregressive LLM generation. We enable all available optimizations, including \texttt{torch.compile}, and chunked prefills for vLLM v1; DeepSpeed ZeRO Stage 3, Unsloth and FlashAttention for LlamaFactory.
We allocated 4, 8, and 16 A100 GPUs for the three models respectively.

\Cref{fig:end-to-end} shows the results. The separate approach uses tensor parallelism (degree 1, 2, 4) for the three models respectively, creating four pipelines in total. We explored various strategies to distribute resources between inference and finetuning tasks. Configurations like 25\% vLLM - 75\% LlamaFactory allocate 25\% of resources (i.e., 1 pipeline) to inference, 75\% (i.e., 3 pipelines) to finetuning.

Configurations with fewer inference pipelines (25\%-50\% vLLM) handle only lightweight workloads, achieving limited inference throughput and SLO attainment under more intensive workloads. Dedicating more pipelines to inference (75\% vLLM) maintains high SLO attainment but reduces finetuning throughput.

\Sys maintains near-optimal SLO attainment without compromising finetuning performance by prioritizing latency-sensitive inference and opportunistically maximizing finetuning tokens.

Across all three models, \sys matches the 75\% vLLM - 25\% LlamaFactory configuration in inference SLO attainment (at or above 90\% even at 20 req/s) and inference throughput, while dramatically improving finetuning throughput. Specifically, under heavy inference loads (20 req/s), \sys sustains finetuning throughputs of 7.2K, 2.2K and 2.2K tokens/s for LLaMA-3.1-8B, Qwen-2.5-14B, and Qwen-2.5-32B respectively, compared to only 3.8K, 1.0K, and 0.5K tokens/s in the 75\% vLLM - 25\% LlamaFactory setup—i.e. a \textbf{1.9$\times$-4.8$\times$} improvement. Under light inference loads (4.0 req/s), \sys sustains finetuning throughputs of 9.4K, 3.7K, and 3.2K tokens/s for the same models, translating to a \textbf{2.5$\times$-6.8$\times$} improvement over the 75\% vLLM - 25\% LlamaFactory setup.

\subsection{GPU Scheduling}

This section compares \sys's co-serving mechanism with two commonly used GPU scheduling strategies: {\em temporal sharing} and {\em spatial sharing}. 
Temporal sharing interleaves the execution of inference and finetuning tasks over time, as illustrated in \Cref{fig:baseline}. Spatial sharing simultaneously launches inference and finetuning computations using separate CUDA resources. Both strategies are implemented on top of \sys by replacing the co-serving mechanism with their respective scheduling strategies. To ensure a fair comparison, all memory optimizations are enabled in each baseline. 

\Cref{fig:temporal-spatial-sharing} shows the results. For temporal sharing, interleaving one inference iteration with one finetuning iteration would violate the SLOs for nearly all inference requests.
This is because each inference iteration is expected to complete within tens of milliseconds, whereas a finetuning iteration requires several seconds. To mitigate this issue, in the temporal sharing experiments, we interleave each finetuning iteration with $n$ inference iterations, where $n$ is the inference frequency shown in the figure.

While adopting a low inference frequency in temporal sharing (i.e., frequency = 64) maximizes finetuning throughput, it adversely impacts the SLO attainment and throughput for inference requests.
Employing a frequency of 128 enables temporal sharing to match the inference throughput and SLO attainment rate of co-serving. However, it also reduces finetuning throughput by 0.57x-0.86$\times$ compared to co-serving.

Finally, we also implemented a dynamic temporal sharing baseline (see \Cref{sec:dts-appendix}), where the interleaving frequency is determined dynamically based on queue lengths, batch sizes, arrival rates, and completion rates.
Dynamic temporal sharing outperforms fixed-frequency temporal sharing by adapting to workload conditions, maintaining SLO attainment above 90\% in most scenarios, and achieving impressive inference throughputs of 5.4K, 4.9K, and 6.6K tokens/s for the three models under heavy loads (20 req/s). However, it still lags behind co-serving, whose finetuning throughput is \textbf{1.0–1.7$\times$} higher, and shows instability under the heaviest loads as SLO attainment drops significantly for the 32B model.

Spatial sharing allocates separate GPU resources for inference and finetuning, achieving comparable finetuning throughput to co-serving. However, it remains suboptimal in SLO attainment under heavy inference workloads due to the interference between inference and finetuning tasks.

\begin{figure}[t]
    \centering
    \begin{subfigure}[b]{\linewidth}
      \includegraphics[width=\linewidth]{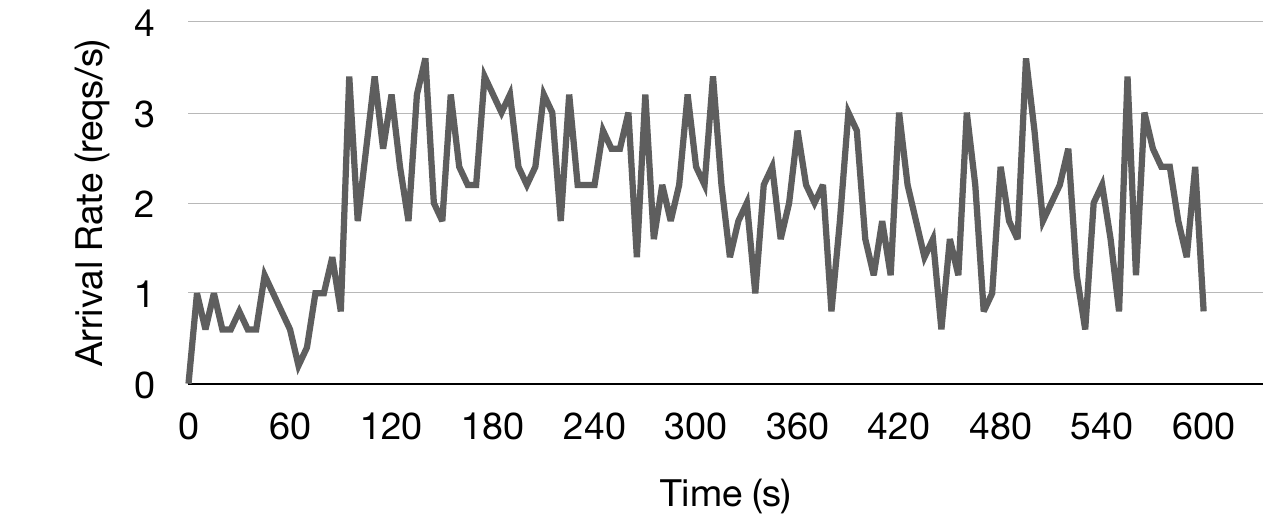}
      \caption{Arrival Inference Requests}
    \end{subfigure}\\ 
    \centering
    \begin{subfigure}[b]{\linewidth}
      \centering
      \includegraphics[width=\linewidth]{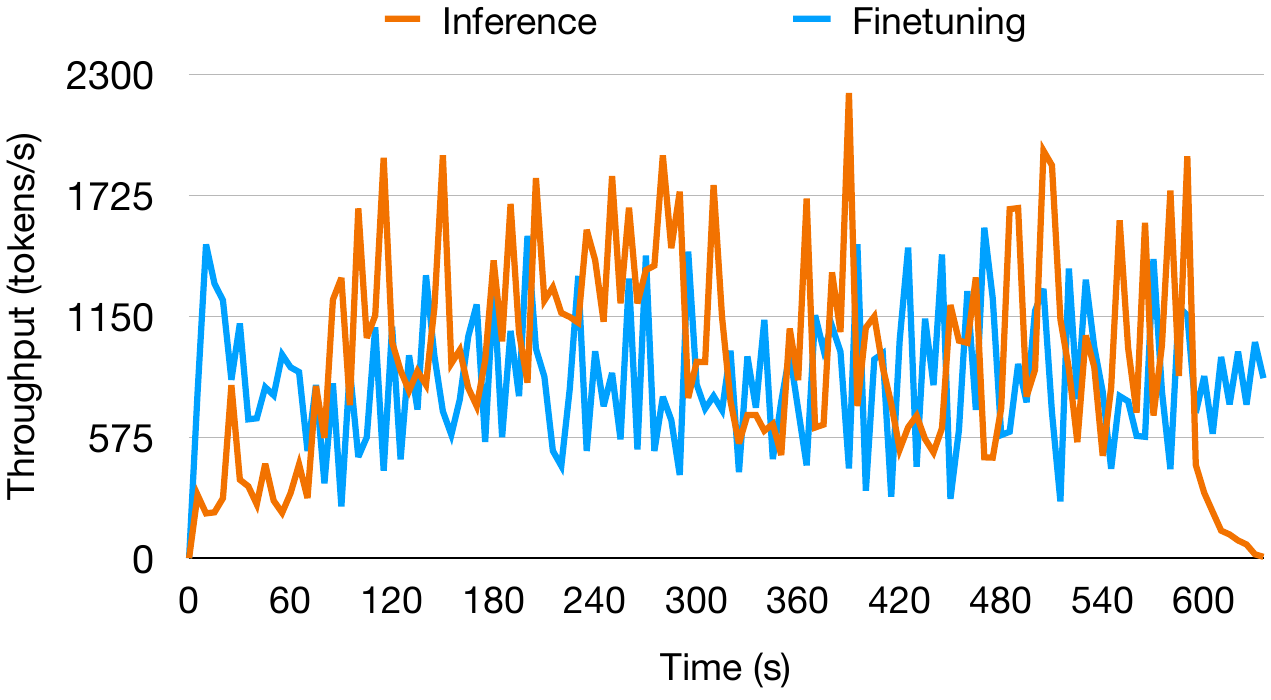}
      \caption{System throughput}
    \end{subfigure}
  \caption{Case study of \sys's system throughput for fluctuating inference workload.}
  \label{fig:case_study}
\end{figure}

\subsection{Case Study}
\label{sec:exp:dynamic}
This case study demonstrates \sys's ability to dynamically adapt to fluctuating inference workloads in real-time. We evaluate \sys using a 10-min interval of the BurstGPT trace with Qwen-2.5-14B model. To ensure compatibility with our experiment environments, we replayed this trace segment and re-scaled its arrival intensity, as previous works have done~\cite{285173,258862,ali2022optimizing,miao2023spotserve}. In the experiment, each inference request and inference request are sampled from the ShareGPT dataset and the Sky-T1 dataset respectively.

In our case study, as shown in Figure~\ref{fig:case_study}, we observed that the arrival rate of the inference requests initially increased to a peak level after around 90 seconds, and then gradually decreased with some peaks. \sys could automatically detect the fluctuations in the workload and improve the ratio of inference tokens (vs finetuning tokens) in each iteration's batch. This significantly increased inference throughput from a few hundreds to 2.25K.

\subsection{Memory Optimization}
\label{eval:memory_optimization}
\begin{figure}[t]
    \centering
    \includegraphics[width=0.85\linewidth]{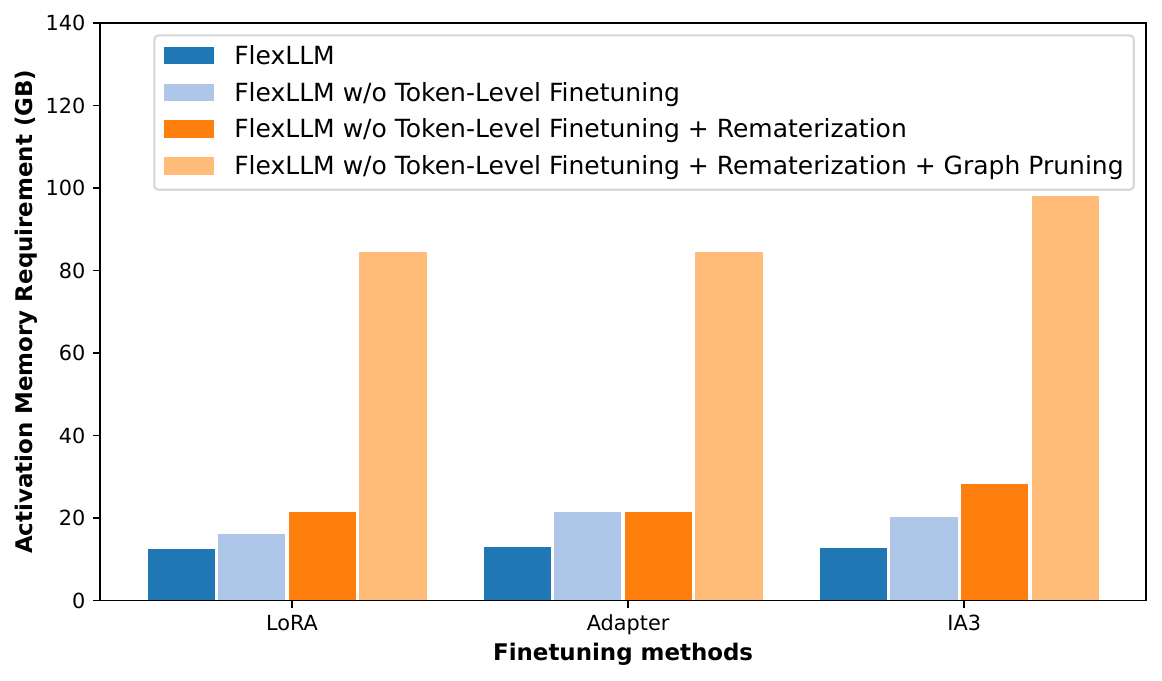}
    \caption{Ablation study of memory optimizations.}
    \label{fig:memory_ablation}
\end{figure}

\sys introduces a series of memory optimizations to reduce the memory overhead of \peft finetuning and allow co-serving finetuning with inference workloads.
To understand the effectiveness of \sys memory optimizations, we perform an ablation study that incrementally turns off these optimizations (graph pruning, rematerialization, and token-level finetuning) and measures the memory requirements. 
\Cref{fig:memory_ablation} shows the activation memory requirements of \sys for different finetuning methods on a 70B LLM and a sequence length of 1024. \sys saves $85\%$-$87\%$ memory requirements for activations compared to existing approaches. The major improvement comes from graph pruning. With graph pruning only, \sys achieves $71\%$-$74\%$ activation memory overhead reduction. Rematerialization and token-level finetuning further reduce memory overhead by $0$-$8\%$ and $4\%$-$10\%$, respectively. These memory optimizations are highly effective in practice, enabling \sys to retain sufficient memory for the inference requests' KV cache, ensuring eviction rates of 0\% in most cases and peaking at only 1.2\% for the largest model (Qwen2.5-32B) under the heaviest loads (see \Cref{sec:coserving_evictions}).

\section{Related Work}
\label{sec:related}

\paragraph{ML serving and inference systems.} Numerous ML serving systems~\cite{258862,gunasekaran2022cocktail,lee2018pretzel,romero2021infaas,crankshaw2017clipper} address challenges like latency prediction, scalability, swapping~\cite{bai2020pipeswitch}, preemption~\cite{choi2020prema}, workload estimation~\cite{285173}, and cost-effectiveness~\cite{zhang2019mark}. Systems like Nexus~\cite{shen2019nexus} and Gpulet~\cite{choi2022serving} focus on batching and GPU virtualization for small models. 
Recent LLM serving systems include vLLM~\cite{10.1145/3600006.3613165}, Sarathi~\cite{agrawal2023sarathi}, SGLang~\cite{sglang}, AlpaServe~\cite{li2023alpaserve}, Punica~\cite{chen2023punica}, and S-LoRA~\cite{sheng2023s}, which improve throughput through batching and specialized PEFT kernels. Aqua~\cite{aqua2024} introduces fine-grained fair scheduling among multiple inference workloads.
Speculative decoding~\cite{specsampling, miao2023specinfer, suffixdecoding} can reduce serving latency and improve SLO attainment~\cite{adaserve} and goodput~\cite{turbospec}; combining it with co-serving is future work.

\paragraph{GPU resource sharing for ML.} GPU resource sharing focuses on efficient allocation among multiple ML users. Lyra~\cite{li2023lyra} allocates dedicated resources with dynamic adjustment. Gandiva~\cite{xiao2018gandiva} and Antman~\cite{xiao2020antman} use time-slicing for better cluster utilization. GSLICE~\cite{dhakal2020gslice} develops adaptive allocation schemes for different SLO constraints. Although these approaches can not directly solve the resource utilization problem in PEFT service, we believe GPU resource sharing will be increasingly important as GPUs become more powerful.
\section{Limitations and Future Work}
While \sys demonstrates significant improvements in GPU utilization and finetuning throughput, we identify several promising directions for future work. First, extending our implementation beyond offline finetuning scenarios where sequences are processed individually (batch size 1), to reinforcement-learning-based approaches like RLHF~\cite{sheng2024hybridflow, Stiennon2020} and Group Relative Policy Optimization (GRPO)~\cite{guo2025deepseek, MultiObjGRPO, DAPO} presents an exciting opportunity, as our token-level co-serving architecture naturally fits these methods where auto-regressive generation and gradient updates are tightly coupled. Second, developing adaptive strategies for workloads with consistently high inference demand could further expand co-serving opportunities. Finally, advancing beyond our current simple scheduling heuristics to more sophisticated algorithms that optimize for task priorities, deadlines, and QoS requirements could unlock additional system efficiency gains.
\section{Conclusion}
This paper proposes \sys, a system for co-serving parameter-efficient finetuning and inference of large language models. We observe the distinct GPU resource utilization features between finetuning and inference workloads and propose a holistic PEFT serving system that supports to co-serve finetuning requests without affecting inference latency.

\section*{Acknowledgement}
We would like to thank the anonymous reviewers and our shepherd, Hitesh Ballani, for their valuable comments and suggestions.
This research is partially supported by NSF awards CNS-2211882 and CNS-2239351, a Sloan research fellowship, and research awards from Amazon, Cisco, Google, Meta, NVIDIA, Oracle, Qualcomm, and Samsung.

\newpage

\balance
\bibliography{bibliography}
\bibliographystyle{plain}

\newpage

\appendix

\section{Dynamic Temporal Sharing}
\label{sec:dts-appendix}

Dynamic Temporal Sharing (DTS) is an adaptive baseline that dynamically adjusts the frequency of switching between inference and finetuning phases based on real-time system conditions. The algorithm (\Cref{alg:dts}) continuously monitors queue lengths, batch sizes, arrival rates, and completion rates to compute a multi-dimensional pressure metric comprising queue pressure ($\text{avg\_queue}/20.0$), spike pressure ($\text{max\_queue}/25.0$), and backlog pressure ($(\text{arrival\_rate} - \text{completion\_rate})/8.0$). Based on the total pressure, DTS adaptively selects finetuning intervals between 64-512 steps using pressure thresholds (pressure $\leq 0.8$ for minimum frequency, pressure $\geq 2.0$ for maximum frequency) and linear interpolation with a $0.6\times$ scaling factor for intermediate values. To ensure system stability, the algorithm incorporates hysteresis through weighted historical averaging ($(\text{frequency} + 2 \times \text{prev\_frequency})/3$), applies stabilization adjustments to computed frequencies, and implements decision delays where frequency recomputation occurs only every third scheduling decision to prevent oscillatory behavior under varying workloads.

\begin{algorithm}[htbp]
\label{alg:dts}
\caption{Dynamic Temporal Sharing}
\begin{algorithmic}[1]
\Require $q$, $b$, $a$, $c$
\State $Q[]$, $B[]$, $r_a$, $r_c$
\State $s$, $f_p$, $d$

\Procedure{Scheduler\_Step}{$q$, $b$, $a$, $c$}
    \State $r_a \leftarrow r_a + a$
    \State $r_c \leftarrow r_c + c$
    \State $Q$.append($q$)
    \State $B$.append($b$)
    
    \State $s \leftarrow s - 1$
    \If{$s \leq 0$}
        \State $d \leftarrow d + 1$
        \If{$d \geq 3$}
            \State $s \leftarrow$ \Call{Compute\_Next\_Interval}{}
            \State $d \leftarrow 0$
        \Else
            \State $s \leftarrow \min(512, f_p \times 1.1)$
        \EndIf
        \State \Call{Reset\_Stats}{}
        \State \Return True \Comment{switch to finetuning}
    \EndIf
    \State \Return False \Comment{continue inference}
\EndProcedure

\Procedure{Compute\_Next\_Interval}{}
    \If{$Q$.empty()}
        \State \Return 64
    \EndIf
    
    \State $\bar{q} \leftarrow$ mean($Q$)
    \State $q_{max} \leftarrow$ max($Q$)
    \State $\bar{b} \leftarrow$ mean($B$)
    \State $\lambda \leftarrow r_a / |Q|$
    \State $\mu \leftarrow r_c / |Q|$
    
    \State $p_q \leftarrow \min(1.0, \bar{q} / 20.0)$
    \State $p_s \leftarrow \min(0.5, q_{max} / 25.0)$
    \State $p_b \leftarrow \max(0.0, (\lambda - \mu) / 8.0)$
    \State $p \leftarrow p_q + p_s + p_b$
    
    \If{$p \leq 0.8$}
        \State $f \leftarrow 64$
    \ElsIf{$p \geq 2.0$}
        \State $f \leftarrow 512$
    \Else
        \State $p_n \leftarrow (p - 0.8) / 1.2$
        \State $f \leftarrow 64 + p_n \times 0.6 \times (512 - 64)$
    \EndIf
    
    \State $f \leftarrow f \times 1.35$ \Comment{stabilization adjustment}
    \State $f_s \leftarrow (f + 2 \times f_p) / 3$
    \State $f_p \leftarrow f_s$
    \State $f_s \leftarrow \max(f_s, 64 + 16)$
    
    \State \Return clamp($f_s$, 64, 512)
\EndProcedure
\end{algorithmic}
\end{algorithm}

\section{Coserving Eviction Rates}
\label{sec:coserving_evictions}
\Cref{tab:coserving_evictions} shows the percentage of inference requests experiencing a KV cache eviction during the evaluation of co-serving under the end-to-end experiment described in \Cref{sec:eval}. Overall, the eviction rates are negliglible. The eviction rate for the Qwen2.5-32B (TP=4) model is 0.29\% for when the inference requests arrive at a rate of 16req/s, and reaches a maximum of 1.20\% when the arrival rate is 20req/s. In all other scenarios, no evictions occurred.

\begin{table}[H]
\centering
\caption{Percentage of requests experiencing an eviction}
\label{tab:coserving_evictions}
\adjustbox{width=\columnwidth,center}{%
\begin{tabular}{|l|c|c|c|c|c|}
\hline
Model & QPS=4 & QPS=8 & QPS=12 & QPS=16 & QPS=20 \\
\hline
Llama-3.1-8B & 0.00\% & 0.00\% & 0.00\% & 0.00\% & 0.00\% \\
Qwen2.5-14B & 0.00\% & 0.00\% & 0.00\% & 0.00\% & 0.00\% \\
Qwen2.5-32B & 0.00\% & 0.00\% & 0.00\% & 0.29\% & 1.20\% \\
\hline
\end{tabular}%
}
\end{table}

\section{Preventing Tenant Interference with Virtual Token Counter}
\label{sec:vtc-integration}

FlexLLM can prevent tenant interference (e.g., noisy neighbor problems) thanks to the Virtual Token Counter (VTC) algorithm~\cite{sheng2024fairness}, which can be combined with our co-serving scheduler. This integration ensures fair resource allocation across tenants while maintaining high GPU utilization.

In multi-tenant PEFT serving, tenants may submit requests at vastly different rates. Without fairness controls, aggressive tenants can monopolize resources and degrade service for others. VTC addresses this by tracking per-tenant service consumption and prioritizing underserved tenants.

Algorithm~\ref{alg:flexllm-vtc} shows the integration of VTC into FlexLLM's token-level scheduling. Key modifications include: (1) per-tenant virtual counters $c_i$ tracking cumulative service, (2) counter lifting when idle tenants rejoin to prevent unfair credit accumulation, (3) fair selection of both inference and finetuning tokens based on minimum counter values, and (4) weighted counter updates reflecting different costs of input ($w_p$), output ($w_q$), and finetuning ($w_r$) tokens.

The integration guarantees bounded fairness across tenants. In the analysis, finetuning requests are treated as a special case of inference requests, allowing us to directly apply results from the Virtual Token Counter analysis~\cite{sheng2024fairness}. Let $W_i(t_1, t_2)$ denote the weighted service received by tenant $i$ and $M$ be the maximum number of tokens. The following lemma and theorems formalize the fairness bounds.

\begin{lemma}
When $Q \neq \emptyset$,
\begin{equation}\label{eq:eq-vtc}
    \max_{i \in Q} c_i - \min_{i \in Q} c_i \le \max(w_p \cdot L_{\text{input}}, \max(w_q, w_r) \cdot M).
\end{equation}
\end{lemma}

\begin{theorem}[Fairness for overloaded tenants]
    For any tenants $f$ and $g$, for any time interval $[t1,t2)$ in which $f$ and $g$ are backlogged, Algorithm~\ref{alg:flexllm-vtc} guarantees
\[
    |W_f(t_1, t_2) - W_g(t_1, t_2)| \leq 2\max(w_p \cdot L_{\text{input}}, \max(w_q, w_r) \cdot M).
\]
\end{theorem}

\begin{theorem}[Fairness for non-overloaded tenants]
    Let $[t_1, t_2)$ be any time interval, $T(t,t2)$ be the total services received for all tenants during $[t,t2)$, $n(t,t2)$ be the number of tenants that have requested services during the interval, and $U$ is the upper bound from Equation~\ref{eq:eq-vtc}.

    Assume a tenant $f$ is not backlogged at $t_1$ and has requested services less than $\frac{T(t,t2)}{n(t,t2)}-5U$ during $[t_1, t_2)$. Then, all of the services requested from $f$ during the interval $[t1,t2)$ will be dispatched.
\end{theorem}

The algorithm remains work-conserving—idle resources are allocated to active tenants—while preventing any single tenant from monopolizing the system. The overhead is minimal: $O(n)$ counters for $n$ tenants, with updates occurring at FlexLLM's existing token-level granularity. This integration preserves FlexLLM's memory optimizations (dependent parallelization, graph pruning) while adding fairness guarantees essential for multi-tenant deployments.

\begin{algorithm}[t]
\caption{FlexLLM with Virtual Token Counter for Fair Co-Serving}
\label{alg:flexllm-vtc}
\begin{algorithmic}[1]
\State \textbf{Input:} Input weight $w_p$, output weight $w_q$, finetuning weight $w_r$, max tokens $M$
\State \textbf{Initialize:} $B \leftarrow \emptyset$, $c_i \leftarrow 0$ for all tenant $i$, $Q \leftarrow \emptyset$
\State
\LeftComment{Monitoring Stream}
\While{True}
    \If{request $r$ from tenant $u$ arrives}
        \If{$\not\exists r' \in Q, tenant(r') = u$}
            \If{$Q = \emptyset$}
                \State $l \gets$ the last tenant left $Q$
                \State $c_u \gets \max\{c_u, c_l\}$
            \Else
                \State $c_u \gets \max\{c_u, \min\{c_i \mid i \in Q\}\}$
            \EndIf
        \EndIf
        \State $Q_u \leftarrow Q_u \cup \{r\}$
    \EndIf
\EndWhile
\State
\LeftComment{Execution Stream}
\While{True}
    \If{can\_add\_new\_requests()}
        \LeftComment{Fair inference selection}
        \While{memory available for inference}
            \State $k \leftarrow \arg\min_{i \in \{tenant(r) \mid r \in Q \text{ is an inference task}\}} c_i$
            \State $r \gets$ the earliest inference request in $Q$ from $k$
            \State Add inference request from $Q_k$ to batch $B$
            \State $c_k \leftarrow c_k + w_p \cdot \text{input\_length}(r)$
        \EndWhile
        
        \LeftComment{Fair finetuning token selection}
        \State $s \leftarrow$ compute\_finetuning\_window()
        \While{$s > 0$ and memory available}
            \State $k \leftarrow \arg\min_{i \in \{tenant(r) \mid r \in Q \text{ is a finetuning task}\}} c_i$
            \State $t \gets \min(s, \text{available})$
            \State Process next $t$ tokens from tenant $k$
            \State $c_k \gets c_k + w_r \cdot t$
            \State $s \leftarrow s - t$
        \EndWhile
    \EndIf
    
    \State Execute decode step for $B$
    \For{each tenant $i$ with tokens generated}
        \State $c_i \leftarrow c_i + w_q \cdot \text{output\_tokens}_i$
    \EndFor
\EndWhile
\end{algorithmic}
\end{algorithm}

\section{Component-wise Memory Breakdown}
\label{sec:memory-breakdown-detailed}

\begin{figure}[p]
    \centering
    \includegraphics[width=\columnwidth]{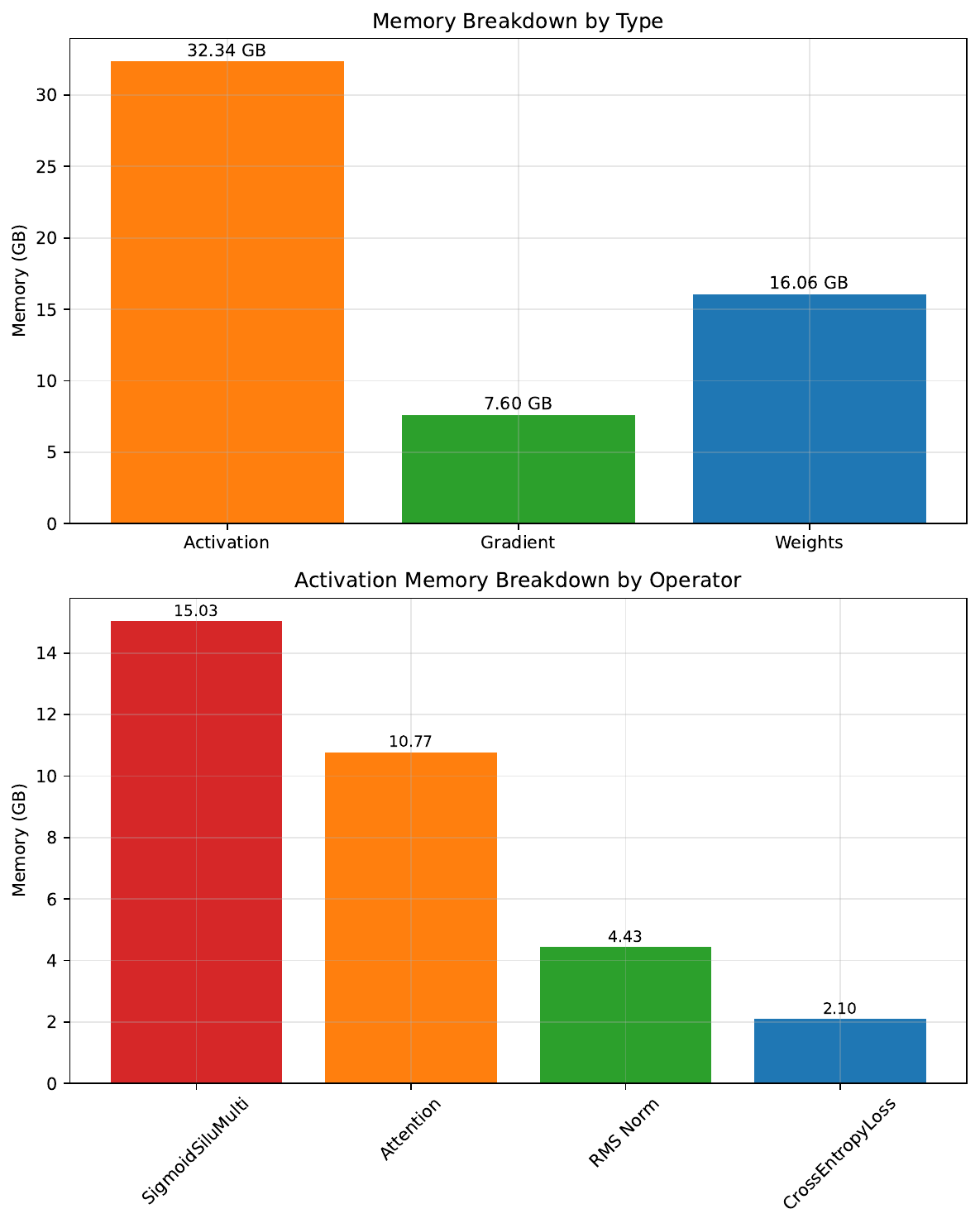}
    \caption{Component-wise memory breakdown for FlexLLM co-serving LLaMA-3.1-8B with LoRA finetuning.}
    \label{fig:memory-breakdown-detailed}
\end{figure}

\Cref{fig:memory-breakdown-detailed} shows the component-wise memory breakdown for FlexLLM co-serving LLaMA-3.1-8B with LoRA rank 16. For finetuning memory management, FlexLLM preallocates a fixed budget sufficient for the largest supported PEFT configuration (maximum rank with all target modules enabled). This budget covers PEFT weights, gradients, optimizer momentum values, and low-rank activations. The static allocation strategy prevents memory fragmentation during dynamic co-serving while ensuring deterministic memory bounds regardless of the specific PEFT configurations being served.

\section{Deployment Considerations}
\label{sec:deployment}

FlexLLM excels in scenarios where its co-serving architecture can maximize resource utilization while maintaining inference performance. \Cref{tab:decision_framework} provides a decision framework to help practitioners identify when FlexLLM delivers optimal benefits.

FlexLLM is particularly well-suited for: (1) inference workloads with moderate burstiness and periods of low utilization, (2) environments with substantial and ongoing finetuning demands, and (3) deployments with moderate SLO requirements (50-100ms TPOT). These conditions allow FlexLLM to fully exploit its co-serving capabilities and memory optimizations.

\begin{table}[t]
\centering
\caption{Decision framework for FlexLLM adoption}
\label{tab:decision_framework}
\adjustbox{width=\columnwidth,center}{%
\small
\begin{tabular}{|l|c|c|}
\hline
\textbf{Scenario} & \textbf{FlexLLM} & \textbf{Separate Clusters} \\
\hline
Bursty inference + high finetuning & \checkmark & \\
Consistent high inference load & & \checkmark \\
Minimal finetuning requirements & & \checkmark \\
Moderate SLOs (50-100ms TPOT) & \checkmark & \\
Strict SLOs (<25ms TPOT) & & \checkmark \\
Cost-sensitive deployments & \checkmark & \\
Operational simplicity priority & & \checkmark \\
\hline
\end{tabular}%
}
\end{table}

FlexLLM's implementation comprises approximately 9K lines of code distributed across compilation, scheduling, memory management, and kernel integration components. The system leverages advanced techniques in GPU programming, distributed systems, and ML compiler optimizations to achieve its performance gains.

Under stricter SLO requirements, FlexLLM's performance characteristics follow predictable queueing theory principles: co-serving introduces service time variance that can affect tail latencies. This trade-off between resource efficiency and latency guarantees becomes more pronounced as SLO targets approach inherent inference latency bounds, making FlexLLM most effective in moderate SLO environments where it can fully utilize its resource optimization capabilities.

\section{Artifact}
The artifact is available at \url{https://github.com/FlexLLM/artifact}. Follow the instructions in the README to build FlexLLM and reproduce the experiments in this paper.

\end{document}